\documentclass[aps,prd,twocolumn,superscriptaddress,preprintnumbers,nofootinbib]{revtex4-1}%
\usepackage{epsf,epsfig}
\usepackage[utf8]{inputenc}
\usepackage{amssymb,amsmath,amsfonts}
\usepackage{color}
\usepackage{tensor}
\usepackage{slashed}
\usepackage{braket}
\usepackage[colorlinks=true]{hyperref}  
\hypersetup{
    bookmarks=true,         
    unicode=false,          
    pdftoolbar=true,        
    pdfmenubar=true,        
    pdffitwindow=false,     
    pdfstartview={FitH},    
    colorlinks=true,       
    linkcolor=magenta, 
    citecolor=blue,        
    filecolor=magenta,      
    urlcolor=cyan           
} 

\newcommand{\bea}{\begin{eqnarray}}
\newcommand{\eea}{\end{eqnarray}}
\newcommand{\ba}{\begin{eqnarray}}
\newcommand{\ea}{\end{eqnarray}}

\newcommand{\beq}{\begin{equation}}
\newcommand{\eeq}{\end{equation}}
\newcommand{\beqa}{\begin{eqnarray}}
\newcommand{\eeqa}{\end{eqnarray}}
\newcommand{\beqar}{\begin{eqnarray*}}
\newcommand{\eeqar}{\end{eqnarray*}}

\newcommand{\ie}{{\it i.e.,}\ }

\newcommand{\E}{\mathcal{E}}

\usepackage{color}

\newcommand{\req}[1]{(\ref{#1})} 

\begin{document}

\title{Extremal Rotating Black Holes in Einsteinian Cubic Gravity}
\author{Pablo A. Cano}
\email{pabloantonio.cano@kuleuven.be}
\affiliation{Instituto de F\'isica Te\'orica UAM/CSIC,
	C/ Nicol\'as Cabrera, 13-15, C.U. Cantoblanco, 28049 Madrid, Spain}
\affiliation{Instituut voor Theoretische Fysica, KU Leuven
	Celestijnenlaan 200D, B-3001 Leuven, Belgium}

\author{David Pere\~niguez}
\email{david.perenniguez@uam.es}
\affiliation{Instituto de F\'isica Te\'orica UAM/CSIC,
	C/ Nicol\'as Cabrera, 13-15, C.U. Cantoblanco, 28049 Madrid, Spain}

\date{October 23, 2019}

\preprint{IFT-UAM/CSIC-19-136}

\begin{abstract} 
We obtain new solutions of Einsteinian cubic gravity coupled to a Maxwell field that describe the near-horizon geometry of charged and rotating black holes. We show that the AdS$_2\times\mathbb{S}^2$ near-horizon geometry of Reissner-Nordstr\"om black holes receives no corrections, but deviations with respect to the extremal Kerr-Newman solution appear as we turn on the angular momentum. We construct the profile of these corrected geometries using both numerical methods and slowly-spinning expansions, but we also find additional solutions that do not reduce to AdS$_2\times\mathbb{S}^2$ geometries in any limit and that do not have a counterpart in Einstein gravity. 
Quite remarkably, we are able to obtain closed-form exact expressions for the area and Wald's entropy of all of these solutions, and using this result, we analyze the phase space of extremal back holes in this theory. To the best of our knowledge, this is the first time the entropy of a rotating black hole in higher-order gravity has been exactly computed.
\end{abstract}

\maketitle

\section{Introduction}
General Relativity describes accurately the dynamics of the gravitational field in the regime of relatively low curvature, but modifications of this theory are expected to appear at high energies. The fact that GR is incompatible with quantum mechanics \cite{tHooft:1974toh,Deser:1974cz,Deser:1974xq} indicates that it should be regarded as an effective theory, presumably arising from an underlying theory of quantum gravity. Independently of what the UV-completion of GR turns out to be, it is broadly accepted that an effective low-energy description of that theory will contain the Einstein-Hilbert action plus an infinite tower of higher-derivative corrections  --- this is, in particular, a definite prediction of String Theory \cite{Gross:1986mw,Gross:1986iv,Grisaru:1986vi,Grisaru:1986px,Bergshoeff:1989de,Green:2003an,Frolov:2001xr}. 
Such corrections modify the behaviour of the gravitational field when the distances involved are of the order of the length scale of new physics. Thus, they become extremely relevant in the very early universe or near black hole singularities, but also at the level of the horizon of small enough black holes. It is therefore an interesting task to determine the properties of the modified black hole solutions, with particular emphasis on the corrections to the thermodynamic quantities, such as entropy and temperature \cite{Myers:1988ze,Wald:1993nt,Iyer:1994ys,Jacobson:1993vj,Cai:2001dz}.


From the point of view of Effective Field Theory (EFT), one should treat the higher-derivative corrections as perturbations over the GR geometry. Obtaining the corrected solutions in this perturbative approach is usually an accessible task; however, perturbative solutions give us very little information. In fact, the perturbative corrections are only valid as long as they remain very small, and many potentially interesting phenomena, that would appear at a non-perturbative level, are lost. For this reason, it is interesting to find exact black hole solutions of higher-order gravity. 

The problem of obtaining exact black hole solutions is, of course, more complicated. Let us consider first the case of spherically symmetric black holes. Until very recently, the only theories in which exact solutions modifying in a non-trivial way the Schwarzschild geometry had been constructed were Lovelock \cite{Lovelock1, Lovelock2,Wheeler:1985nh,Wheeler:1985qd,Boulware:1985wk,Cai:2001dz,Dehghani:2009zzb,deBoer:2009gx,Camanho:2011rj} and Quasi-topological gravities \cite{Quasi,Quasi2,Dehghani:2011vu,Cisterna:2017umf}, both types of theories existing only in $D>4$ dimensions.\footnote{There are theories in which Einstein metrics are exact solutions (\textit{e.g.} if the Lagrangian only contains Ricci curvature \cite{delaCruzDombriz:2009et,Li:2017ncu}), and other that possess ``non-Schwarzschild'' solutions \cite{Lu:2015psa,Lu:2015cqa}. We are not including these in our discussion.} 
The gap in $D=4$ has been recently filled thanks to the construction of a new type of theories with very interesting properties. Known as Generalized Quasi-topological gravities (GQTGs) \cite{Hennigar:2017ego}, these theories allow for simple spherically symmetric black hole solutions whose thermodynamic properties can be studied analytically \cite{Hennigar:2017ego,PabloPablo3,Ahmed:2017jod}. Besides, GQTGs exist in all dimensions (including, in particular, $D=4$) and at all orders in curvature \cite{Bueno:2019ycr}, and very likely they provide a basis to construct the most general EFT for gravity \cite{Bueno:2019ltp}. Spherically symmetric solutions in these theories have been studied at all orders in curvature in $D=4$ \cite{PabloPablo4} and at cubic \cite{Hennigar:2017ego,Mir:2019ecg} and quartic order \cite{Ahmed:2017jod,Mir:2019rik} in various dimensions, and this has allowed us to gain substantial information about spherically symmetric black holes in higher-order gravity. In particular, one of the most remarkable features of these theories is that black holes become stable below certain mass \cite{PabloPablo4}, hence avoiding the complete evaporation in a finite time and the final explosion of black holes. This is analogous to the behaviour of higher-dimensional Lovelock black holes found long ago in Ref.~\cite{Myers:1988ze}.  In this paper we will consider an extension of Einstein gravity containing the simplest non-trivial Generalized Quasi-topological density in $D=4$, which is known as Einsteinian cubic gravity (ECG) \cite{PabloPablo}. This theory was the first member of the GQT class to be discovered and we review some of its properties as well as recent results in Sec.~\ref{sec:ECG}.

Despite the success in the construction of spherically symmetric black holes, a remaining issue in the world of higher-order gravities is to find rotating black hole geometries.\footnote{Let us note that exact rotating black hole solutions have been constructed numerically for some scalar-tensor theories containing higher-curvature terms \cite{Kleihaus:2015aje,Delsate:2018ome}, but not for pure gravity theories.} In fact, exact rotating solutions have not even been found in Lovelock gravity, which is the simplest non-trivial extension of GR that one could consider.\footnote{A honorable exception is the solution found in Ref. \cite{Anabalon:2009kq}, corresponding to a rotating black hole in $D=5$ Gauss-Bonnet gravity at a special point of the parameter space in which there is a unique maximally symmetric solution.} Thus, the question about what a rotating black hole in higher-derivative gravity is like has not been answered yet. However, this is a primordial question, since, after all, realistic black holes will in general possess angular momentum.

The equations of motion for an axisymmetric and stationary metric are far more complicated than those in the spherically symmetric case. Even though we expect some simplification of the equations taking place for GQTGs --- because they do so in the static case ---, obtaining a complete rotating black hole solution would necessarily require a laborious numeric computation. However, there are several limits in which the problem is simplified. On the one hand, one might consider slowly-rotating solutions and stay perturbative in the spin. This has been explored in the case of quadratic \cite{Kim:2007iw} and cubic \cite{Yue:2011et} Lovelock gravity. The case for $D=4$ ECG will be reported in a coming publication \cite{RotECG}. On the other hand, it is possible to study the opposite limit, namely, the case of extremal black holes. In this situation, the horizon is placed at an infinite distance and the near-horizon limit is well-defined, giving rise to a new solution of the gravitational equations. This near-horizon geometry has more symmetries than the global solution, and this enormously simplifies the problem of solving the field equations. We will show in this paper that the equations of motion of ECG reduce in this case to a single second-order ODE. This equation has to be solved numerically, but most remarkably, we will see that it is possible to obtain the exact expressions for the area and entropy of these black holes without using any approximation.
We are not aware that a similar analysis has been performed for other pure-metric higher-order gravities, but let us mention that Ref. \cite{Chen:2018jed} computed the (perturbative) corrections to the near-horizon geometry of extremal Kerr black holes in the case of Einstein-dilaton-Gauss-Bonnet \cite{Kanti:1995vq,Torii:1996yi,Alexeev:1996vs} and dynamical Chern-Simons \cite{Alexander:2009tp} gravities. 

For generality purposes, we will add as well a Maxwell field into the game, which will allow us to study rotating and charged extremal black holes. This will prove to be useful, as AdS$_2\times\mathbb{S}^2$ geometries --- corresponding to non-rotating charged black holes --- are always solutions of higher-order gravities. The rotating black holes can then be studied as a deformation of these geometries, which facilitates the analysis of the solutions. However, we will also show that there are new branches of solutions that do not reduce to AdS$_2\times\mathbb{S}^2$ geometries in any limit. These solutions do not exist in the Einstein gravity limit and, as we will see, they have somewhat exotic properties. 

The paper is organized as follows. We start in Sec.~\ref{sec:ECG} by introducing our theory, corresponding to ECG coupled to a Maxwell field.   In Sec.~\ref{sec:nhg} we write the metric and vector ans\"atze for a rotating near-horizon geometry possessing an $\mathrm{SL}(2,\mathbb{R})\times\mathrm{U}(1)$ isometry group, and we evaluate and partially solve the equations of motion. We reduce the field equations to a single second-order ODE for one variable. Then we discuss the boundary conditions that need to be imposed in order to obtain fully regular solutions. In Sec.~\ref{sec:AdS2} we study in detail the solutions of the previous equation that are smooth deformations of AdS$_2\times\mathbb{S}^2$ geometries. We construct solutions --- both numerically and in the slowly-rotating approximation --- which are labeled by the total charge $Q$ and by a parameter $x_0$ which we argue is related to the spin  $a=J/M$. More interestingly, we find that both the area and the Wald's entropy can be obtained exactly, and we study them as functions of $Q$ and $x_0$. In addition, the physically meaningful relation $S(\mathcal{A},Q)$ is derived and we also study its profile. In Sec.~\ref{sec:add} we analyze the full space of near-horizon geometries, showing that there exists an important degeneracy of solutions. We discuss the properties of the additional branches and comment on the structure of the diagram $S(\mathcal{A},Q)$. Finally, we draw our conclusions in Sec.~\ref{sec:disc}. We also include a number of appendices that support and extend some of the results in the main text.

\section{Einsteinian cubic gravity}\label{sec:ECG}
We consider the following theory
\begin{equation}\label{ECGF2}
S=\frac{1}{16\pi G}\int d^4x\sqrt{|g|}\left\{-2\Lambda+R-\frac{\mu L^4}{8}\mathcal{P}-F_{\mu\nu}F^{\mu\nu}\right\}\, ,
\end{equation}
which consists of the (cosmological) Einstein-Maxwell action --- where $F_{\mu\nu}=2\partial_{[\mu}A_{\nu]}$ --- plus a cubic curvature correction $\mathcal{P}$, the Einsteinian cubic gravity density \cite{PabloPablo}
\begin{align}\label{eq:P}
\mathcal{P}&=12 \tensor{R}{_{\mu}^{\rho}_{\nu}^{\sigma}}\tensor{R}{_{\rho}^{\alpha}_{\sigma}^{\beta}}\tensor{R}{_{\alpha}^{\mu}_{\beta}^{\nu}}+\tensor{R}{_{\mu\nu}^{\rho\sigma}}\tensor{R}{_{\rho\sigma}^{\alpha\beta}}\tensor{R}{_{\alpha\beta}^{\mu\nu}}\\ \notag&-12R_{\mu\nu\rho\sigma}R^{\mu\rho}R^{\nu\sigma}+8\tensor{R}{_{\mu}^{\nu}}\tensor{R}{_{\nu}^{\rho}}\tensor{R}{_{\rho}^{\mu}}.
\end{align}
Also,  $\mu$ is a dimensionless coupling while $L$ is a length scale that determines the distance at which the gravitational interaction is modified.

As stated earlier, $\mathcal{P}$ is the lowest-order non-trivial member of the GQT family of theories in $D=4$.  On a historical note, this theory was first identified by the special form of its linearized equations on maximally symmetric backgrounds, which turn out to be of second order in any dimension \cite{PabloPablo}. Afterwards, the simple form of spherically symmetric black hole solutions in this theory was noticed \cite{Hennigar:2016gkm,PabloPablo2}, and this triggered the construction of the GQT class of theories \cite{Hennigar:2017ego,PabloPablo3,Ahmed:2017jod}.  By now,  many other aspects of ECG have been explored, including the characterization of observational deviations with respect to GR \cite{Hennigar:2018hza,Poshteh:2018wqy}, holographic applications \cite{Dey:2016pei,ECGholo,Bueno:2018yzo}, inflationary cosmologies \cite{Arciniega:2018fxj,Cisterna:2018tgx,Arciniega:2018tnn}\footnote{In the cosmological context, the solutions appearing in Refs. \cite{Arciniega:2018fxj,Cisterna:2018tgx,Arciniega:2018tnn} were constructed in a modified cubic theory that takes the form $\mathcal{P}-8\mathcal{C}$, where $\mathcal{P}$ is the ECG term --- see \req{eq:P} --- and $\mathcal{C}$ is a cubic piece that makes no contribution on spherically symmetric backgrounds \cite{Hennigar:2016gkm}. Thus, that combination produces the same black hole solutions as the ones constructed for ECG. We have checked that $\mathcal{C}$ is irrelevant for our present setup too. } and other types of solutions \cite{Feng:2017tev,Bueno:2018uoy,Mehdizadeh:2019qvc}.

Up to the six-derivative level, $\mathcal{P}$ represents the leading parity-preserving higher-derivative correction to the Einstein-Hilbert action \cite{Bueno:2019ltp}. However, when a Maxwell field is included, there are other terms that we could add at this order. Schematically, these would be of the form $F^4$, $R F^2$, $F^6$, $R F^4$, $R^2 F^2$. Nevertheless, it is not our intention to study the most general correction to extremal Kerr-Newman geometries. Instead, we focus on the theory above because it will allow as to perform many explicit computations that are practically unaccessible for other higher-derivative theories. 

The equations of motion of \req{ECGF2} read
\begin{align}\label{EOMs}
\mathcal{E}_{\mu\nu}&=T_{\mu\nu}\, ,\\ \notag
\nabla_{\mu}F^{\mu\nu}&=0\, ,
\end{align}
where the gravitational tensor $\mathcal{E}_{\mu\nu}$ and the energy-momentum tensor $T_{\mu\nu}$ are given by 
\begin{align}\label{eq:fe}
\mathcal{E}_{\mu\nu}&=G_{\mu\nu}+\Lambda g_{\mu\nu}\\ \notag
&-\frac{\mu L^4}{8} \left(P_{\mu\sigma\rho\lambda}R_{\nu}\,^{\sigma \rho\lambda}-\frac{\mathcal{P}}{2}g_{\mu\nu}+2\nabla^{\alpha}\nabla^{\beta}P_{\mu\alpha\nu\beta}\right)\, ,\\
T_{\mu\nu}&=2F_{\mu\alpha}\tensor{F}{_{\nu}^{\alpha}}-\frac{1}{2} g_{\mu\nu} F_{\alpha\beta}F^{\alpha\beta}\, ,
\end{align}
and where
\begin{equation}
\begin{aligned}
\label{eq:PECG}
\tensor{P}{_{\mu\nu}^{\alpha\beta}}&=36\tensor{R}{_{[\mu|\sigma}^{[\alpha|}_{\rho}} \tensor{R}{_{|\nu]}^{\sigma|\beta]}^{\rho}}+3\tensor{R}{_{\mu\nu}^{\sigma\rho}}\tensor{R}{_{\sigma\rho}^{\alpha\beta}}\\ 
&-12 \tensor{R}{_{[\mu}^{[\alpha}}\tensor{R}{_{\nu]}^{\beta]}}-24R^{\sigma\rho}\tensor{R}{_{\sigma[\mu|\rho}^{[\alpha}}\tensor{\delta}{_{|\nu]}^{\beta]}}\\ 
&+24\tensor{R}{_{\sigma}^{[\alpha|}}\tensor{R}{^{\sigma}_{[\mu}}\tensor{\delta}{_{\nu]}^{|\beta]}}\, .
\end{aligned}
\end{equation}

\section{Near-horizon geometries}\label{sec:nhg}
Near-horizon geometries of extremal rotating black holes possess an isometry group $\mathrm{SL}(2,\mathbb{R})\times\mathrm{U}(1)$, and a general ansatz for this type of metrics can be written as \cite{Astefanesei:2006dd}
\begin{align}
ds^2&=(x^2+n^2)\left(-r^2dt^2+\frac{dr^2}{r^2}\right)\\ \notag&+\frac{dx^2}{f(x)}+N(x)^2f(x)(d\psi-2nr dt)^2\, ,
\end{align}
which depends on two functions $f(x)$ and $N(x)$ and on one constant $n$. In addition, we consider a vector field of the following form
\begin{equation}
A=h(x)(d\psi-2nr dt)\, ,
\end{equation}
which depends on another function $h(x)$. Then, we have to insert this ansatz in the equations of motion \req{EOMs} and solve them.  Due to the symmetries of the ansatz, one can check that the only independent components of the Einstein's equations are those corresponding to $\mu\nu=xx$ and $\mu\nu=\psi\psi$ --- the rest are related to them by the Bianchi identities. Thus, we only need to solve those equations together with the Maxwell equation.

An important observation is that these equations allow for solutions that have $N(x)=1$. The reason is that, when evaluated on $N(x)=1$, the components of the gravitational tensor --- which we show in Appendix \ref{app:eoms} --- become proportional, namely
\begin{equation}\label{eq:Nconst}
\mathcal{E}_{\psi\psi}\Big|_{N(x)=1}=f(x)^2\mathcal{E}_{xx}\Big|_{N(x)=1}\, ,
\end{equation}
and the same property holds for the Maxwell energy-momentum tensor $T_{\mu\nu}$. In general, higher-derivative gravities do not satisfy the condition \req{eq:Nconst}, meaning that these theories do not allow for solutions with constant $N(x)$. In turn, it is quite remarkable that this property holds for ECG. As we are going to see, this represents a drastic simplification of the equations of motion.  Let us also note at this point that, besides the solutions with $N(x)=1$, there can be other solutions. In fact, Einstein gravity allows for solutions with non-constant $N(x)$, but these turn out to be singular, and only the solutions with $N(x)=1$ represent the near-horizon geometry of extremal Kerr-Newman black holes. In the same way, ECG will presumably allow as well for this type of singular solutions when $N(x)$ is non-constant. Thus, from now on we set $N(x)=1$.

Now, we can evaluate Maxwell's equation, which turns out to be independent of $f(x)$:
\begin{equation}
d\star F=\left[\left(h'(x)(x^2+n^2)\right)'+\frac{4n^2h(x)}{x^2+n^2}\right]dt\wedge dt\wedge dx=0\, ,
\end{equation}
where the prime denotes derivation with respect to $x$. The general solution of this equation reads
\begin{equation}
h(x)=\frac{a(x^2-n^2)}{x^2+n^2}+\frac{2bnx}{x^2+n^2}\, ,
\end{equation}
where $a$ and $b$ are two integration constants that are related to the electric and magnetic charges.  Thus, at this point we have reduced the problem to solving one equation for $f(x)$, namely $\mathcal{E}_{xx}=T_{xx}$. However, before going into the resolution of this equation, let us massage a bit the solution in its current form. Let us note that the coordinate $x$ is compact and it will range within two values $x_0>0$ and $-x_0$. These values are determined by the vanishing of the function $f(x)$ --- which is assumed to be even --- at those points: $f(x_0)=f(-x_0)=0$. Also, let us introduce the quantity
\begin{equation}
\omega\equiv -\frac{f'(x_0)}{2}=\frac{f'(-x_0)}{2}>0\, .
\end{equation}
Then, observe that in order to avoid a conical singularity at $x=\pm x_0$ --- these points will correspond to the poles of the horizon --- the coordinate $\psi$ must have period $2\pi/\omega$. Using these results, we can already compute the electric and magnetic charges even if we do not know explicitly the function $f(x)$. In Planck units, these charges read
\begin{align}
q&=\frac{1}{4\pi}\int\star F
=\frac{2anx_0}{\omega(x_0^2+n^2)}\, ,\\
p&=\frac{1}{4\pi}\int F
=\frac{2bnx_0}{\omega(x_0^2+n^2)}\, ,
\end{align}
where the integration is performed on any surface of constant $t$ and $r$. We note that these are the actual values of the charges that we would obtain in a global solution containing an asymptotic region. Let us finally exchange $x$ and $\psi$ in terms of two new coordinates
\begin{align}
x&=x_0y\, ,\quad y\in [-1,1]\, ,\\
\psi&=\frac{\phi}{\omega}\, ,\quad \phi\in [0,2\pi)\, ,\\
\end{align}
and let us introduce the function
\begin{equation}
g(y)=\frac{f(y x_0)}{x_0^2}\, .
\end{equation}
In this way, we rewrite our solution in the following form
\begin{align}\label{solmassaged}
ds^2&=(y^2x_0^2+n^2)\left(-r^2dt^2+\frac{dr^2}{r^2}\right)\\&\notag+\frac{dy^2}{g(y)}+\frac{x_0^2}{\omega^2}g(y)(d\phi-2\omega nr dt)^2\, ,\\
A&=\frac{x_0^2+n^2}{y^2x_0^2+n^2}\left[q\frac{y^2x_0^2-n^2}{2nx_0}+py\right](d\phi-2\omega nr dt)\, ,
\end{align}
and by construction, $g(y)$ satisfies
\begin{equation}\label{conditions}
g(1)=0\, ,\quad g'(1)=-\frac{2\omega}{x_0}\, .
\end{equation}
Let us finally evaluate the remaining equation, which in the new coordinates is $\E_{yy}=T_{yy}$. On the one hand, we have
\begin{equation}
T_{yy}=\frac{\omega^2 \left(x_0^2+n^2\right)^2Q^2}{x_0^2g(y) \left(n^2+x_0^2y^2\right)^2}\, ,
\end{equation}
where $Q^2=q^2+p^2$. 
On the other hand, $\E_{yy}$ takes the form of a total derivative, namely
\begin{equation}
\E_{yy}=\frac{y^2}{(y^2x_0^2+n^2)^2g(y)}\frac{d}{dy}\E(g,g',g'';y)\, ,
\end{equation}
where 
\begin{widetext}
\begin{equation}
\begin{aligned}
\E(g,g',g'';y)&=-\frac{n^2}{y}+y x_0^2+g \left(\frac{n^2 x_0^2}{y}+y x_0^4\right)+\Lambda 
   \left(-\frac{n^4}{y}+2 n^2 y x_0^2+\frac{1}{3} y^3 x_0^4\right)\\&+L^4 \mu  \Bigg[\frac{3
   g^3 n^2 x_0^6 \left(n^2-9 y^2 x_0^2\right)}{y \left(n^2+y^2
   x_0^2\right){}^3}+\left(-\frac{3 g^2 x_0^6 \left(-17 n^2+y^2 x_0^2\right)}{2
   \left(n^2+y^2 x_0^2\right){}^2}-\frac{3 g x_0^4}{2 \left(n^2+y^2 x_0^2\right)}\right)
   g'\\&+\left(-\frac{3 x_0^2}{4 y}-\frac{3 g n^2 x_0^4}{2 n^2 y+2 y^3 x_0^2}\right)
   \left(g'\right)^2+\frac{1}{4} x_0^4 \left(g'\right)^3+g\left(\frac{3 x_0^2}{2
   y}+\frac{3 g x_0^4 \left(-4 n^2+y^2 x_0^2\right)}{2 y \left(n^2+y^2
   x_0^2\right)}-\frac{3}{4} x_0^4 g'\right) g''\Bigg]\, .
\end{aligned}
\end{equation}
\end{widetext}
Hence, integrating both sides of the equation we obtain
\begin{equation}\label{mastereq}
\E(g,g',g'';y)=-\frac{\omega^2 \left(x_0^2+n^2\right)^2Q^2}{x_0^2 y}+N\, ,
\end{equation}
where $N$ is an integration constant. Thus, we have reduced the equations of motion to a single ODE of second order for $g(y)$. 

Our task now is to solve the previous equation in order to obtain near-horizon geometries. So far, we have included a non-vanishing cosmological constant for generality, but for the sake of simplicity we set $\Lambda=0$ from now on.  The case of $\Lambda\neq 0$ is briefly discussed in Appendix \ref{app:lambda}.

\subsection{Einstein gravity}
Let us first of all check that we recover the near-horizon geometry of extremal Kerr-Newman black holes when we set $\mu=0$. In that case, Eq.~\req{mastereq} is simply algebraic and we obtain the solution straightforwardly, 
\begin{equation}
g(y)=\frac{n^2-Q^2 
   \left(n^2+x_0^2\right){}^2\omega ^2/x_0^2+N y-x_0^2 y^2}{x_0^2 \left(n^2+x_0^2 y^2\right)}\, .
\end{equation}
We can see that the parameter $N$ breaks the symmetry $y\leftrightarrow -y$ of the solution that we assumed in identifying the charges $q, p$. More importantly, when $N$ is present (and $x_0\neq0$), there is necessarily a conical singularity at one of the poles of the horizon (where $g$ vanishes), because the slope of $g$ will be different in each one. In fact, $N$ is the NUT charge, and it is known that NUT-charged, rotating black holes present this type of conical singularities at the horizon \cite{Ghezelbash:2007kw}. In order to avoid these problems, we set $N=0$. In that case, $g(y)$ is even, and we have to impose the conditions \req{conditions}, which are going to fix several relations between the parameters of the solution. We find
\begin{equation}
n=\sqrt{Q^2+x_0^2}\, ,\quad \omega=\frac{x_0}{Q^2+2x_0^2}\, ,
\end{equation}
and after simplifying we obtain
\begin{equation}\label{NHEKN}
g(y)=\frac{1-y^2}{Q^2+x_0^2(1+y^2)}\, .
\end{equation}
We see that this is the near-horizon geometry of extremal Kerr-Newman black holes (NHEKN) \cite{Bardeen:1999px}, where $x_0$ is nothing but the angular momentum per mass $x_0=a$. Likewise, $n=M$ is the total mass and $\omega$ is the angular velocity of the horizon. In addition, we can compute the area, which reads
\begin{equation}
\mathcal{A}=\frac{4\pi x_0}{\omega}=4\pi (Q^2+2x_0^2)\, .
\end{equation}
For $x_0=0$ we recover AdS$_2\times\mathbb{S}^2$, which is the near-horizon geometry of extremal Reissner-Nordstrom black holes. 

\subsection{Einsteinian cubic gravity}
Let us now consider a non-vanishing $\mu$.
In analogy to the Einstein gravity case, we set the NUT charge to zero, $N=0$, in order to avoid conical singularities. Now, once the corrections are included, the equation \req{mastereq}  becomes of second order and we need to impose appropriate boundary conditions in order to solve it. We warn that the constraints \req{conditions} are not really boundary conditions: they are restrictions to the parameters of the solution. Instead, the boundary conditions we will impose are the following: (1) the solution is even, and this is equivalent to asking $g'(0)=0$. (2) The solution is regular at $y=\pm 1$, \ie it is analytic at those points. Therefore, according to \req{conditions}, the solution should have a Taylor expansion near $y=1$ of the form
\begin{equation}\label{gexp}
g(y)=-\frac{2\omega}{x_0}(y-1)+\sum_{k=2}^{\infty}g_k (y-1)^k\, ,
\end{equation}
for some coefficients $g_k$. When this expansion is inserted in \req{mastereq} we can Taylor-expand the equation as well, obtaining the following series
\begin{equation}
y\E(g,g',g'';y)+\frac{\omega^2 \left(x_0^2+n^2\right)^2Q^2}{x_0^2}=\sum_{k=0}^{\infty}C_k (y-1)^k\, .
\end{equation}
Thus, all the coefficients $C_k$ must vanish and this gives us a series of equations for the parameters of the solution. Remarkably, the first two equations $C_0$ and $C_1$ are independent of the $g_k$, and instead they provide two relations between $x_0$, $n$, $\omega$ and $Q$:
\begin{align}\label{eq:constr1}
x_0^2-n^2+\frac{Q^2 \omega ^2
   \left(n^2+x_0^2\right){}^2}{x_0^2}-\mu  L^4 \omega ^2 \left(2 x_0 \omega +3\right)&=0\, , \\ \notag
 \left(n^2+x_0^2\right) \left(n^2 \omega +x_0^2 \omega -x_0\right)\hskip3.3cm& \\ \notag
 + \mu  L^4 \omega ^2 \left(-5 n^2 \omega +x_0^2 \omega +3 x_0\right)
   &=0\, .
\end{align}
We have seen that in Einstein gravity $x_0$ is identified with the angular momentum per mass, $a$, while in turn $n$ is the mass and $\omega$ is the angular velocity. We cannot expect that the same identifications work for higher-curvature gravity, and, since we lack the asymptotic region, we cannot correctly identify these quantities. Nevertheless, since $x_0$ controls the degree of non-sphericity of the solution, we do expect that there will be a monotonous relation between this parameter and the angular momentum --- we recall that this parameter enters in the metric as $ds^2=(x_0^2y^2+n^2)ds^2_{{\text AdS}_2}+\ldots$. Hence, it seems reasonable to use $x_0$ and the charge $Q$ to label our solutions. Then, the equations \req{eq:constr1} provide us with the values of $n(x_0, Q)$ and $\omega(x_0, Q)$. It is worth emphasizing that such equations are exact; we have implemented no approximation in our approach. Besides, this allows us to compute the area of these black holes even if we do not know $g$ explicitly, since it is given by
\begin{equation}\label{eq:area}
\mathcal{A}=\frac{4\pi x_0}{\omega}\, .
\end{equation}
Then, once the parameters $n$ and $\omega$ (or alternatively $\mathcal{A}$) are determined, we can solve the rest of the equations $C_2=0$, $C_3=0$, etc. It tuns out that these equations fix all the coefficients $g_{k>3}$ in \req{gexp} in terms of $g_2$, which is the only free parameter. Thus, we find that there is only a one-parameter family of solutions that are regular at the pole $y=1$, which means that regularity is in fact fixing one integration constant. Now, the remaining parameter $g_2$ is determined by the condition that $g$ be an even function, which is equivalent to asking $g'(0)=0$. Thus, we have a well-defined boundary problem, which at most will possess a discrete number of solutions.

\section{The AdS$_2\times\mathbb{S}^2$ branch}\label{sec:AdS2}
Let us summarize our findings so far. Our near-horizon geometries are labelled by two parameters which we can choose to be $Q$ and $x_0$. Imposing regularity of the solution at $y=\pm1$ yields the equations \req{eq:constr1}, whose solutions give the possible values of $n$ and $\omega$.  Finally, the differential equation \req{mastereq} must be solved imposing the regularity condition \req{gexp} and $g'(0)=0$. As we will see later, the equations \req{eq:constr1} have more than one solution for fixed $Q$ and $x_0$, which leads to an important degeneracy of near-horizon geometries that have the same $Q$ and $x_0$. However, it turns out that there is only one branch of solutions that are smoothly connected to an AdS$_2\times\mathbb{S}^2$ geometry in the limit of $x_0\rightarrow0$. In this section we will focus our attention on those solutions.

Let us first solve the equations \req{eq:constr1} when $x_0<<Q$ by assuming a series expansion of the form $\omega=\sum_{n}\omega_k x_0^k$, $n^2=\sum_{k}c_k x_0^k$. We find the following solution
\begin{equation}
\begin{aligned}\label{soladsnw}
n^2=&Q^2+x_0^2 \left(1+\frac{\mu  L^4}{Q^4}\right)+\mathcal{O}(x_0^4)\, ,\\
\omega=&\frac{x_0}{Q^2}+\frac{x_0^3}{Q^4}\left(-2+\frac{\mu L^4}{Q^4}\right)+\mathcal{O}(x_0^5)\, ,
\end{aligned}
\end{equation}
where the higher-order terms can be easily computed as well and we show few of them in Appendix \ref{app:termo}.  Now, let us also assume a series expansion of the metric function $g(y)$, so that
\begin{equation}\label{expansiong}
g(y)=\sum_{k=0}^{\infty}x_0^{2k}g_k(y)\, . 
\end{equation}
Plugging this expansion together with \req{soladsnw} in the equation \req{mastereq} we find the equation satisfied by every component $g_k(y)$. The leading term $g_0$--- which is the only one that survives in the limit $x_0\rightarrow 0$ --- satisfies the following equation
\begin{equation}\label{eq:g0eq}
-1+y^2+g_0 Q^2+\frac{3 L^4 \mu }{4}\left(\frac{4}{Q^4}- \left(g_0'\right)^2+2g_0
    g_0''\right)=0\, .
\end{equation}
We can see that a solution of this equation fulfilling the appropriate boundary conditions is given by
\begin{equation}\label{eq:g0sol}
g_0(y)=\frac{1-y^2}{Q^2}\, .
\end{equation}
Thus, in the limit $x_0\rightarrow 0$ the metric \req{solmassaged} becomes
\begin{equation}
ds^2=Q^2\left(-r^2dt^2+\frac{dr^2}{r^2}\right)+Q^2\left(\frac{dy^2}{1-y^2}+(1-y^2)d\phi^2\right)\, ,
\end{equation}
which corresponds to an AdS$_2\times\mathbb{S}^2$ geometry. In fact, this is the near-horizon geometry of extremal Reissner-Nordstrom black holes, and, as we can see, it possesses no corrections. Thus, this is an exact solution of ECG for any value of $\mu$. Let us then consider the effect of rotation by assuming a finite $x_0$. Analyzing the equations for the following terms, $g_k(y)$, we see that they all allow for a solution which is a polynomial in $y$, and that this solution is the only one that satisfies the boundary conditions. For instance, up to quadratic order in $x_0$ we have
\begin{widetext}
\begin{align}\label{gsmallx0}
g(y)=(1-y^2)\Bigg[\frac{1}{Q^2}
-x_0^2 \frac{\left(Q^8-3\mu L^4 Q^4  +9 \mu^2L^8 +y^2 \left(Q^8-16 \mu L^4 Q^4
   \right)\right)}{Q^8 \left(Q^4-9 L^4 \mu \right)}+\ldots\Bigg]
\end{align}
\end{widetext}
and more terms are shown in the Appendix \ref{app:sol}.  A few comments are in order. First, let us remark that this is a perturbative expansion in $x_0$, but it is exact in $\mu$. Second, we observe that if we put $\mu=0$ in the expression above we get $g(y)=(1-y^2)(Q^{-2}-x_0^2Q^{-4}(1+y^2)+\ldots)$, which coincides with the perturbative expansion of the NHEKN solution \req{NHEKN}, and the same holds for the higher-order terms that we show in the appendix. Therefore, these solutions in principle approach the NHEKN one when $\mu \rightarrow 0$, or more precisely, when $Q>>\mu^{1/4}L$, \textit{i.e.}, when the size of the black hole is much larger than the length scale of the corrections. However, there is a subtlety: we observe that the $\mathcal{O}(x_0^2)$ term (and also all the higher-order ones) diverges for $Q^4=9\mu L^4$. In general, we observe that all the terms of order greater or equal than $2n$ diverge for $Q^4=3((n+1)^2-1)\mu L^4$. This implies that, when $Q$ crosses one of these values, the solution changes discontinuously, and near those critical values we seem to find no solution. Therefore, as we increase $Q$ and $x_0$, the solution will approach the NHEKN one, but it will make it in a non-continuous way.
\begin{figure}[t!]
\includegraphics[scale=0.45]{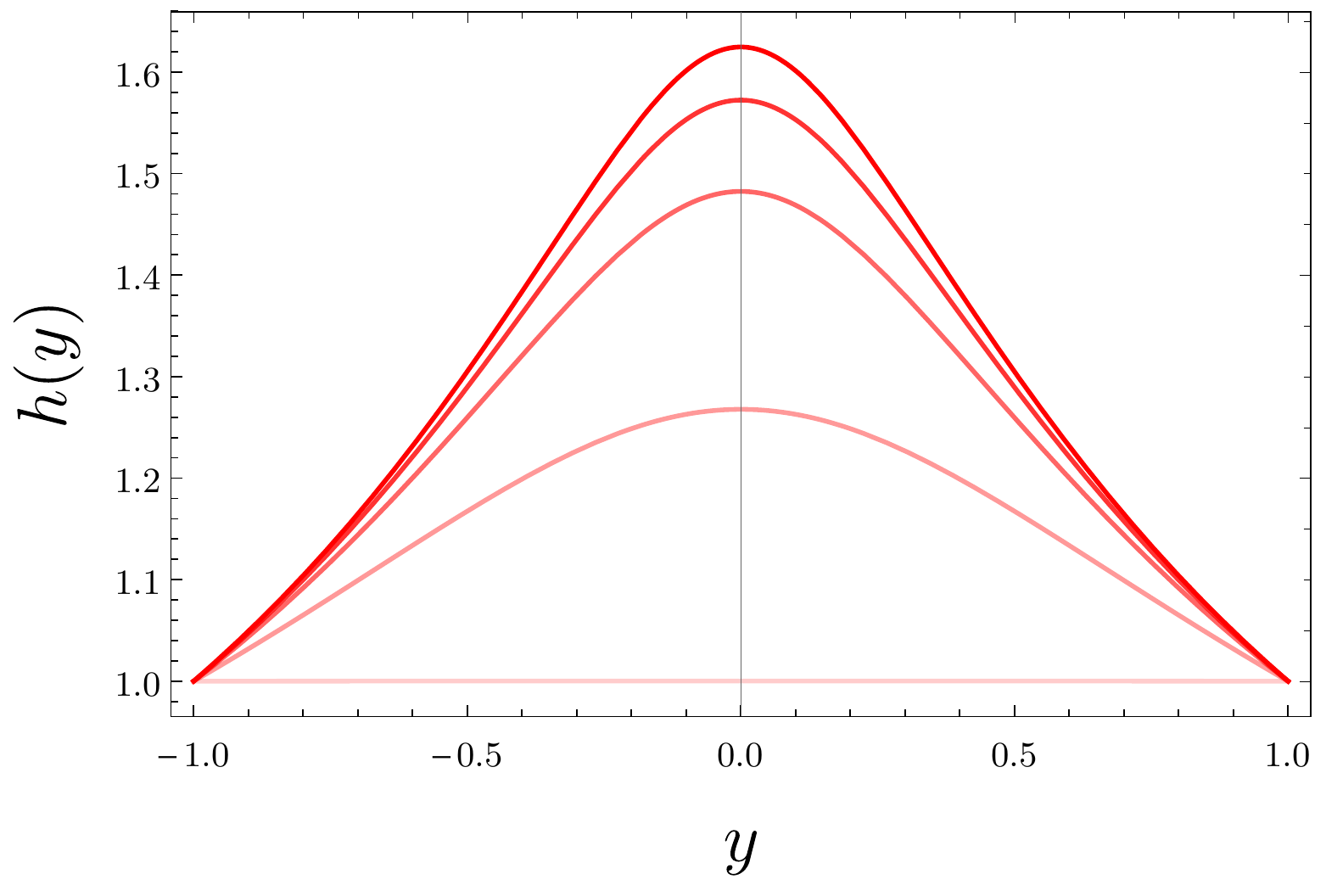}
\includegraphics[scale=0.45]{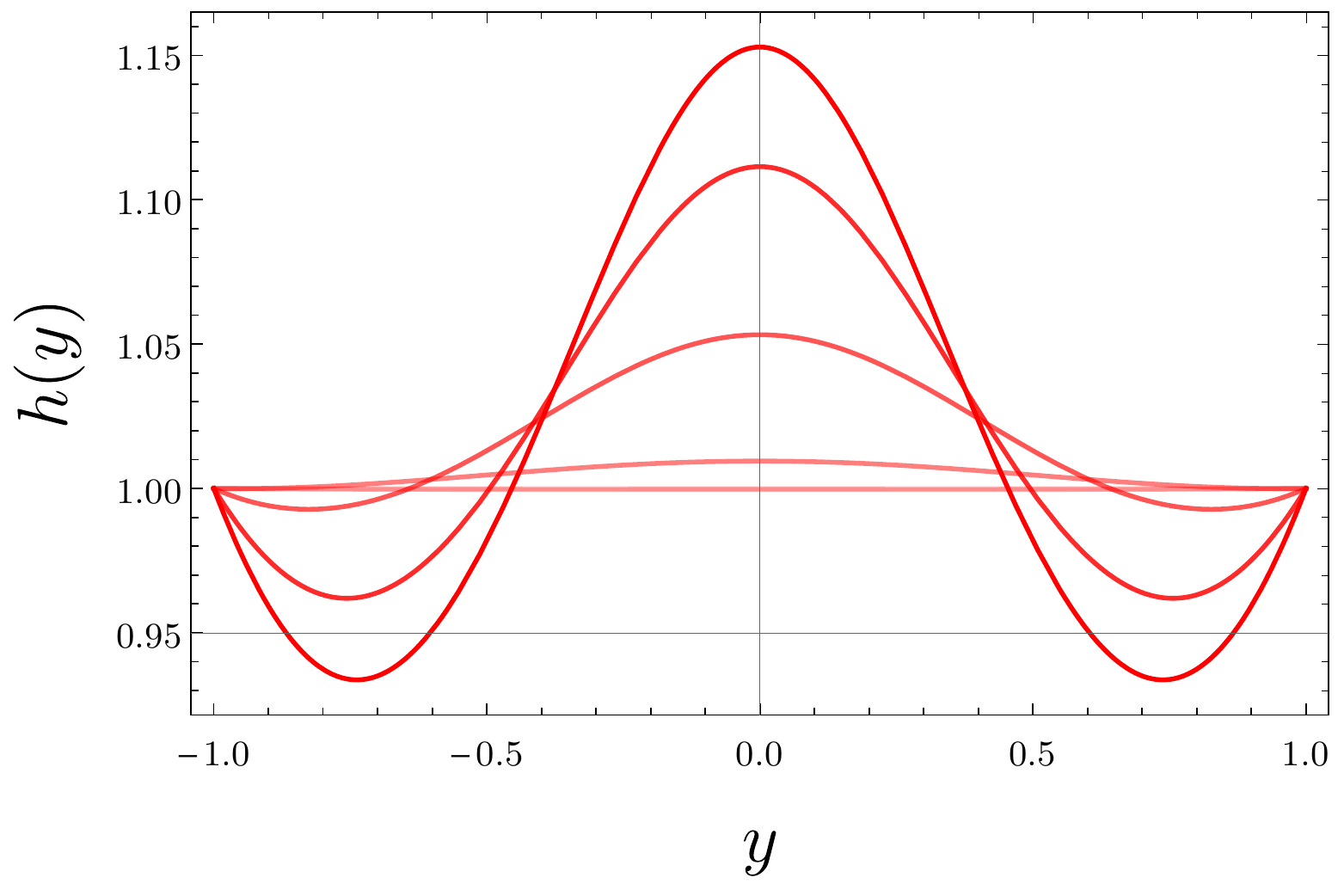}
\includegraphics[scale=0.45]{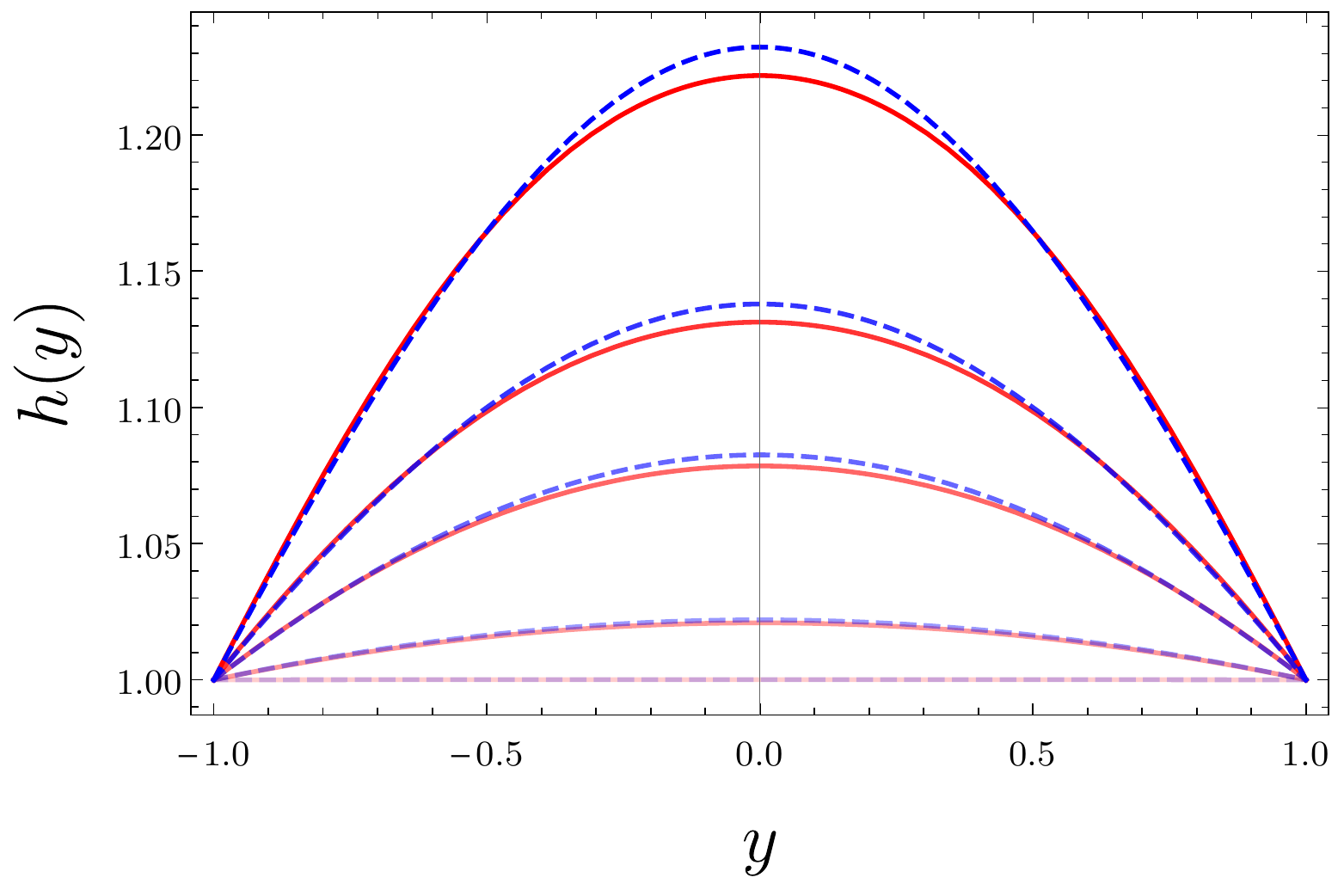}
\caption{Profile of the solution for various values of $Q$ and $x_0$. We show the quantity $h(y)\equiv\frac{x_0g(y)}{\omega(1-y^2)}$, which measures the non-sphericity of the solution (for $\mathbb{S}^2$ this quantity is constant). Top: we show the solution for $Q^4=3\mu L^4$ and $x_0=0, 0.3Q, 0.35Q, 0.36Q, 0.364Q$. Middle: $Q^4=15\mu L^4$ and $x_0=0, 0.1Q, 0.2Q, 0.3Q, 0.34Q$. We observe that the profile is very different in both cases, because we have passed the critical value of $Q^4=9\mu L^4$. Bottom: $Q^4=150\mu L^4$ and $x_0=0, 0.15Q, 0.3Q, 0.4Q, 0.55Q$. The size of the black hole is larger and the solution becomes more similar to the NHEKN one, shown in blue dashed lines for comparison. }
\label{fig:sol}
\end{figure}

\begin{figure}[ht]
\includegraphics[scale=0.4]{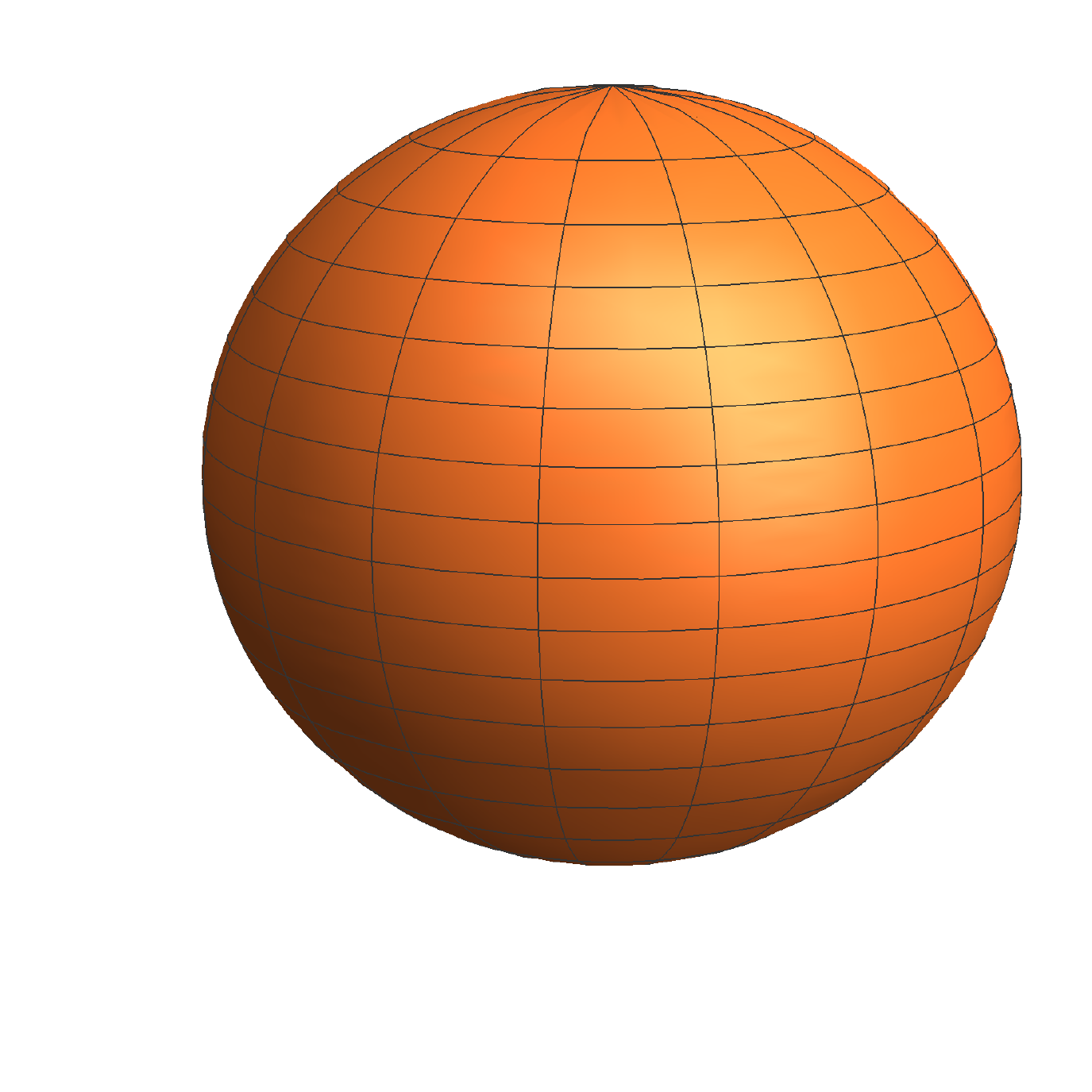}
\includegraphics[scale=0.4]{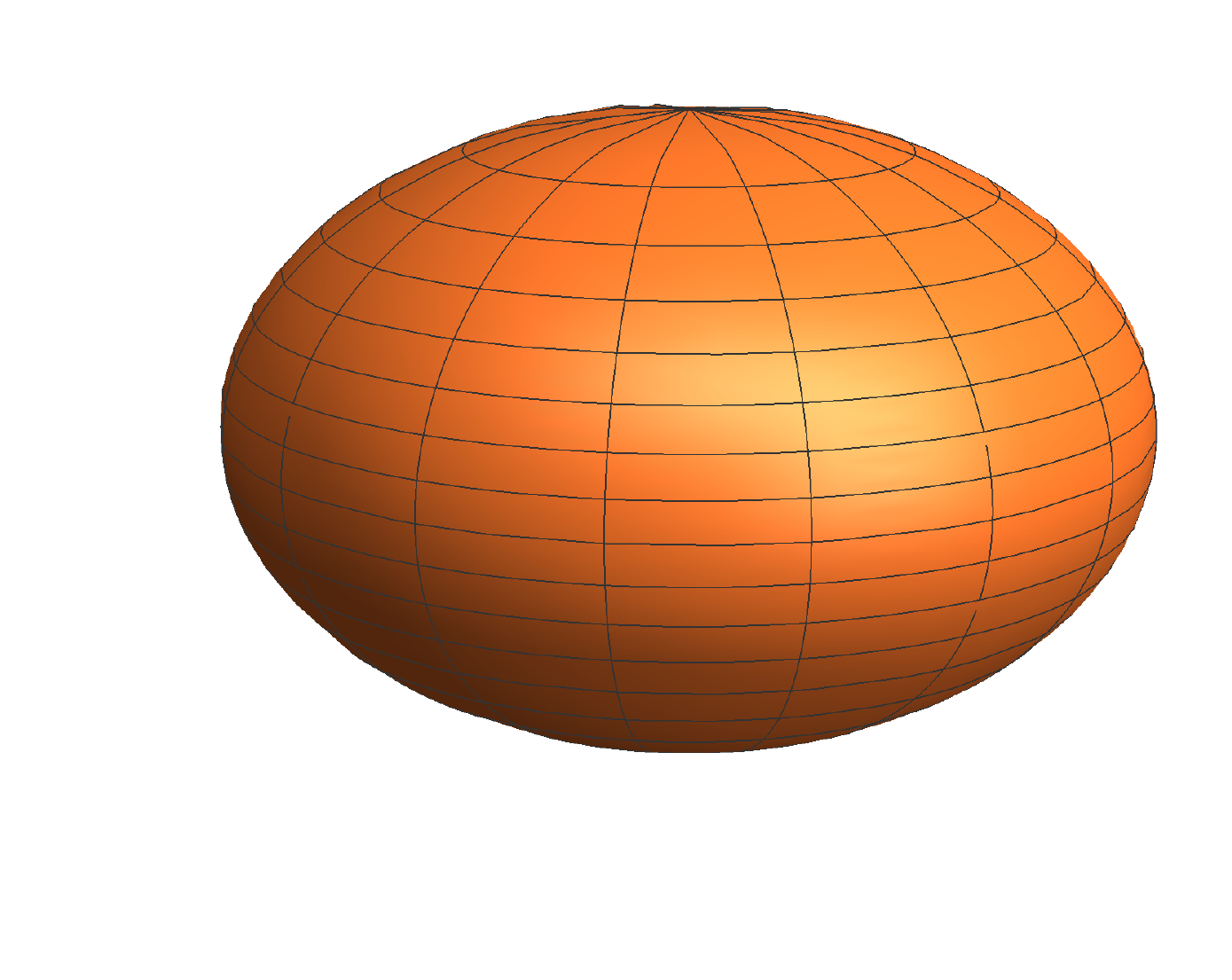}
\includegraphics[scale=0.4]{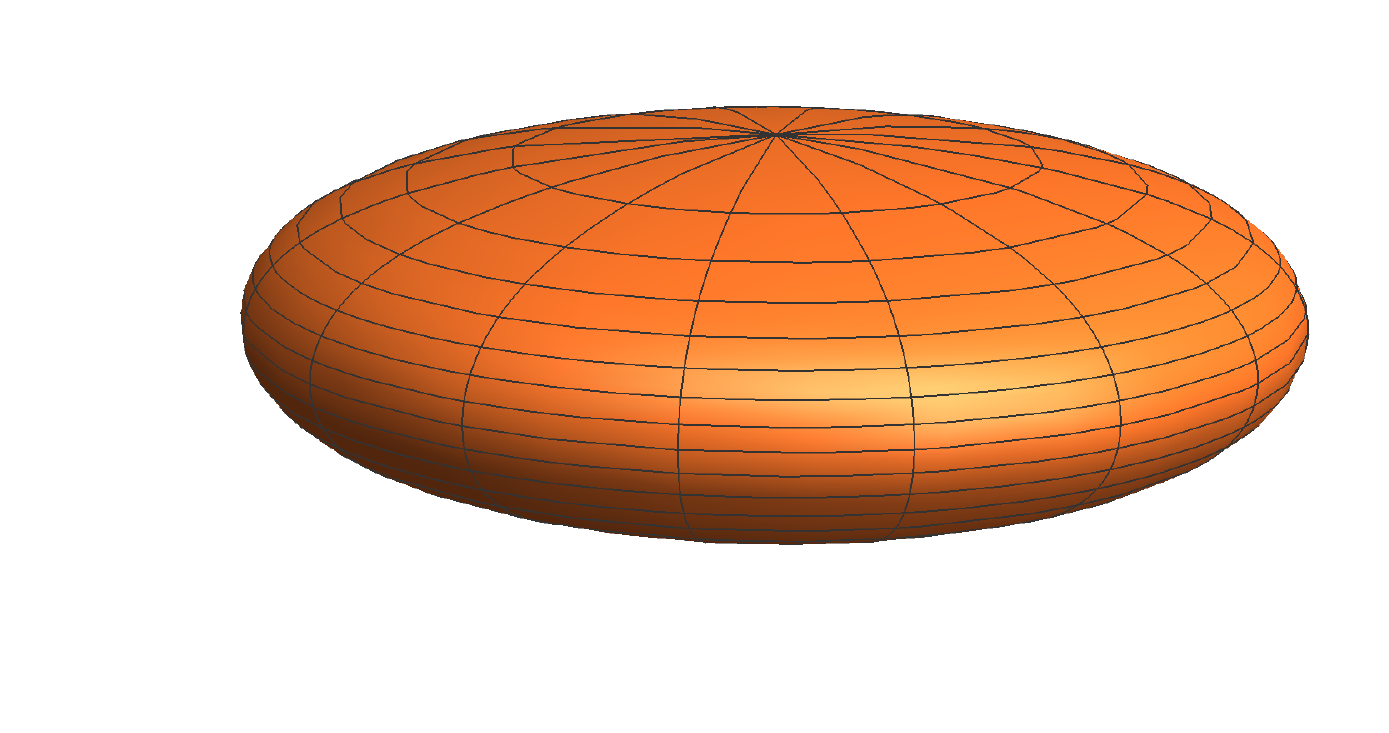}
\caption{Isometric embedding of the horizon in $\mathbb{E}^3$ for the black holes with $Q^4=3\mu L^4$ and $x_0=0$, $x_0=0.3Q$ and $x_0=0.364 Q$.}
\label{fig:embed}
\end{figure}
 This is best understood  by constructing the non-perturbative numerical solutions. We show several of them in Fig.~\ref{fig:sol}, where we represent the function $h(y)\equiv\frac{x_0g(y)}{\omega(1-y^2)}$, which allows for a more direct comparison between the different curves. We have checked that, when $x_0$ is small enough, the slowly-rotating expansion \req{gsmallx0} gives a very good approximation to the numerical curves. Looking at Fig.~\ref{fig:sol} we observe that, indeed, the profile of the solution is quite different for distinct values of $Q$, but eventually it becomes similar to the NHEKN one for large black holes. In addition, in Fig.~\ref{fig:embed} we show the embedding of the black horizon in Euclidean space for some of these solutions.

One important drawback, though, is that we do not seem to find solutions when $x_0/Q$ is large. As we can see, Eq.~\req{mastereq} becomes singular at the points in which $\frac{3 x_0^2}{2
   }+\frac{3 g x_0^4 \left(-4 n^2+y^2 x_0^2\right)}{2  \left(n^2+y^2
   x_0^2\right)}-\frac{3}{4} x_0^4 y g'=0$, which implies that the coefficient of $g''$ vanishes. This only happens when the ratio $x_0/Q$ is large enough. For example, if we evaluate the previous expression for NHEKN geometries and we ask that it does not vanish at any point, we must impose $x_0/Q<1/\sqrt{3}$. Now, if that quantity vanishes, the solution will typically become singular at that point, unless we fix a regularity boundary condition there. But in that case, we cannot impose the boundary conditions of regularity at $y=\pm1$ and that $g'(0)=0$. Hence, we find that, even in the regime where the corrections are small, the equation \req{mastereq} has no regular solutions correcting the NHEKN geometry for $x_0/Q$ large.  
   
In addition, our numerical exploration indicates the existence of an important multiplicity of solutions even when the boundary conditions are fixed.  This is, once we have solved \req{eq:constr1} and found $n(x_0, Q)$, $\omega(x_0, Q)$, the equation \req{mastereq} seems to have different solutions that differ on the profile of $g(y)$. This already happens in the $x_0\rightarrow 0$ equation \req{eq:g0eq}, which possesses other solutions than \req{eq:g0sol} satisfying $g(\pm 1)=0$, $g'(\pm1)=\mp 2/Q^2$. These do not need to be similar to the NHEKN geometry even when $\mu$ is small, and in general they will possess a different domain of existence from the solutions considered in the preceding paragraph. In any case, all of these solutions are characterized by the same set of parameters $x_0$, $Q$, $n$, $\omega$, so they share a number of common properties. 

In order to simplify the discussion, in the next subsection we will remain agnostic about the existence or non-existence of solutions of Eq. \req{mastereq}. Providing some solution exist, we are going to see that the area and entropy can be obtained exactly without knowing the profile of $g(y)$. 

\subsection{Area and entropy}
\begin{figure}[ht]
\includegraphics[width=\columnwidth]{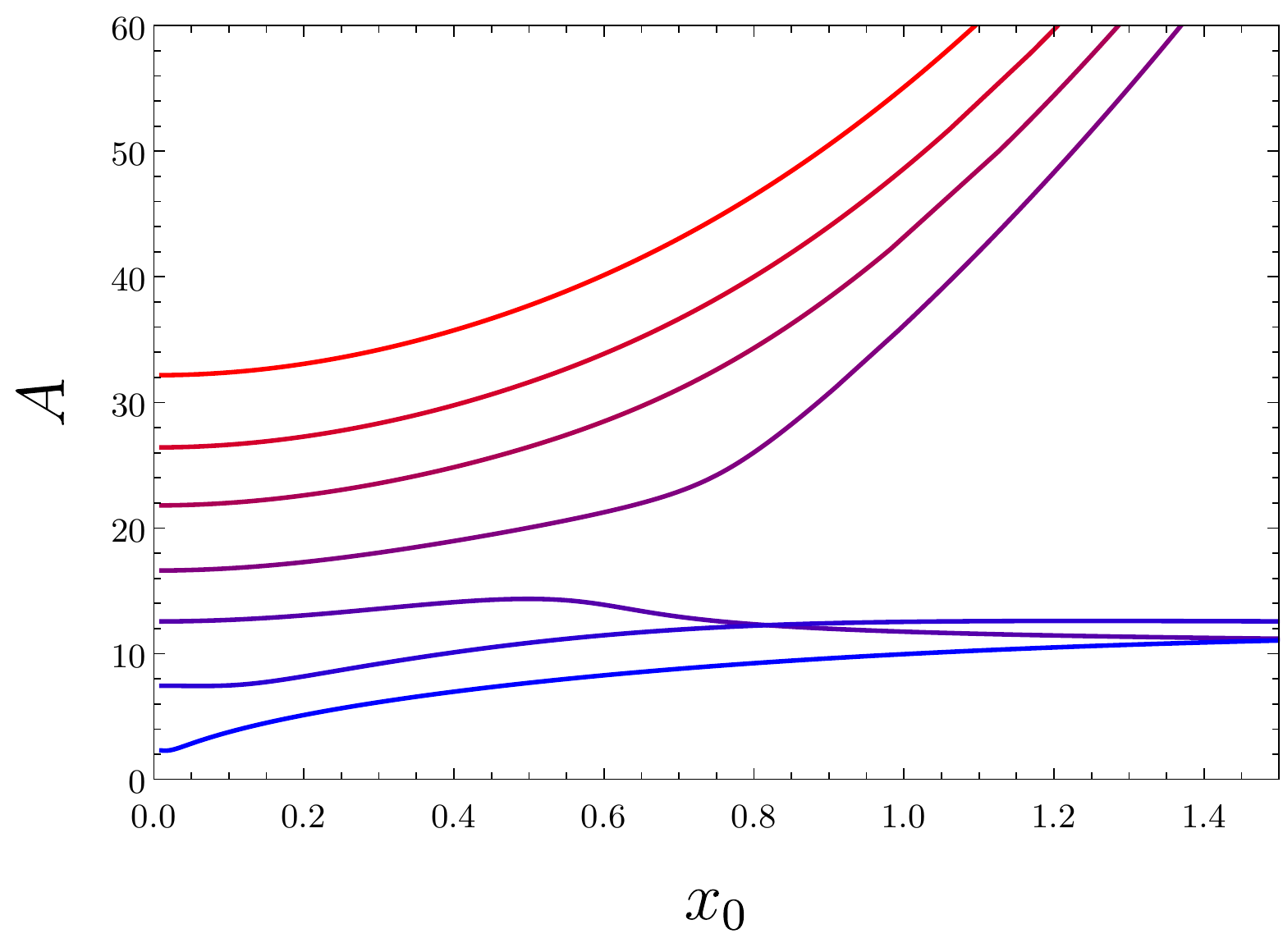}
\caption{Area of black holes that are smooth deformations of AdS$_2\times\mathbb{S}^2$ geometries as a function of $x_0$ for various values of $Q$. From blue to red we have $Q=0.43, 0.77, 1, 1.15, 3^{1/4}, 1.45, 1.6$. We work in units such that $\mu L^4=1$. For large enough $Q$, the curves tend to the Einstein gravity values in both limits $x_0\rightarrow 0,\infty$, but when $Q$ is too small the area tends to a constant value for $x_0\rightarrow\infty$.}
\label{fig:area}
\end{figure}

As we have seen, it is possible to solve the equation \req{mastereq} either perturbatively in $x_0$ or numerically. Nevertheless, there are some properties of these near-horizon geometries that we can compute exactly. One of them is the area, which is given by \req{eq:area}. Then, using Eqs.~\req{eq:constr1} one is able to obtain the area as a function of $x_0$ and $Q$. The relation $\mathcal{A}(x_0,Q)$ for several values of $Q$ is shown in Fig.~\ref{fig:area}.  Near $x_0=0$, one can use the expansions \req{soladsnw} in order to obtain the approximation
\begin{equation}
\frac{\mathcal{A}}{4\pi}=Q^2+ x_0^2 \left(2-\frac{\mu  L^4}{Q^4}\right)+\frac{12 \mu  L^4 x_0^4}{Q^{10}} \left(\mu  L^4-Q^4\right)+\ldots\, ,
\end{equation}
which is valid as long as $x_0<<Q$. Thus, for $x_0\rightarrow 0$, the area reduces to the corresponding value of extremal Reissner-Nordstrom black holes, but looking at Fig.~\ref{fig:area} we see that an interesting behaviour takes place when we increase $x_0$. If the charge is large enough, the corresponding curve differs slightly from the value in Einstein gravity for intermediate values of $x_0$, but for large $x_0$ one recovers again the extremal Kerr-Newman result $\mathcal{A}\sim 4 \pi (Q^2+ 2x_0^2)$. On the other hand, if the charge is too small --- the threshold value is 
\begin{equation}
Q_{\rm thr}\approx 1.13\mu^{1/4} L
\end{equation} 
--- the curve does not approach the Einstein gravity result, and instead we see that $\mathcal{A}$ tends to a constant for $x_0\rightarrow\infty$.  This represents an exotic solution that does not exist in Einstein gravity, and it satisfies
\begin{equation}\label{eq:x0large}
\mathcal{A}=4\pi\alpha ,\quad n^2=\frac{x_0^2 \left(2 \alpha+Q^2\right)}{5 Q^2}\, ,\quad \text{when}\,\,\, x_0\rightarrow\infty\, ,
\end{equation}
where $\alpha$ is a constant determined from the equation
\begin{equation}
2\alpha^3+12\alpha^2 Q^2+18 \alpha Q^4-25 \mu  L^4 Q^2=0\, .
\end{equation}


On the other hand, near-horizon geometries allow us to compute the entropy of black holes, even if we do not know the behavior in the asymptotic region, thanks to Wald's entropy formula \cite{Wald:1993nt,Iyer:1994ys,Jacobson:1993vj}, which reads\footnote{For Lagrangians containing covariant derivatives of the Riemann tensor, the partial derivative of the Lagrangian should be replaced by the Euler-Lagrange derivative of the gravitational Lagrangian as if the Riemann tensor were an independent variable, this is
\begin{equation}
\frac{\delta \mathcal{L}}{\delta R_{\mu\nu\rho\sigma}}=\frac{\partial \mathcal{L}}{\partial R_{\mu\nu\rho\sigma}}-\nabla_{\alpha}\left(\frac{\partial \mathcal{L}}{\partial \nabla_{\alpha}R_{\mu\nu\rho\sigma}}\right)+\ldots
\end{equation}}
\begin{equation}\label{eq:Wald1}
S=-2\pi \int_{\mathcal{H}} d^2x\sqrt{h} \frac{\partial \mathcal{L}}{\partial R_{\mu\nu\rho\sigma}}\epsilon_{\mu\nu}\epsilon_{\rho\sigma}\, .
\end{equation}
In this expression, the integral is taken over the horizon $\mathcal{H}$, $h$ is the determinant of the induced metric on $\mathcal{H}$ and $\epsilon_{\mu\nu}$ is the binormal of the horizon, normalized as $\epsilon_{\mu\nu}\epsilon^{\mu\nu}=-2$. 
Applying Wald's formula \req{eq:Wald1} to our theory \req{ECGF2}, we get
\begin{equation}\label{eq:Wald2}
S=\frac{1}{4G} \int_{\mathcal{H}} d^2x\sqrt{h}\left[1+\frac{\mu L^4}{16}P_{\mu\nu\alpha\beta}\epsilon^{\mu\nu}\epsilon^{\rho\sigma}\right]\, ,
\end{equation}
where $P_{\mu\nu\alpha\beta}$ is the tensor defined in \req{eq:PECG}. The horizon of the metric  \req{solmassaged} is placed at $r=0$, but the integration can be equivalently performed on any slice of constant $t$ and $r$. The non-vanishing components of the binormal read $\epsilon_{tr}=-\epsilon_{rt}=(y^2x_0^2+n^2)$, so that $P^{\mu\nu\alpha\beta}\epsilon_{\mu\nu}\epsilon_{\rho\sigma}=4(y^2x_0^2+n^2)^2P^{trtr}$. Remarkably, we find that this quantity takes the form of a total derivative,
\begin{align}\label{WaldCorrection}
P^{\mu\nu\alpha\beta}\epsilon_{\mu\nu}\epsilon_{\rho\sigma}&=12 x_0^2\frac{d}{dy} \Bigg(-\frac{4 g^2 n^2 y x_0^2}{\left(n^2+y^2 x_0^2\right){}^3}\\&\notag+\frac{4 g n^2
   g'}{\left(n^2+y^2 x_0^2\right){}^2}+\frac{y \left(g'\right)^2}{n^2+y^2 x_0^2}\Bigg)\, .
\end{align}
Therefore, the integral can be performed without knowing the details of $g(y)$ --- we only require the conditions \req{conditions} --- and the entropy reads
\begin{equation}\label{eq:entro}
S=\frac{\pi x_0}{G\omega}\left[1+\frac{3\mu L^4\omega^2}{n^2+x_0^2}\right]\, .
\end{equation}

\begin{figure}[t!]
\includegraphics[width=\columnwidth]{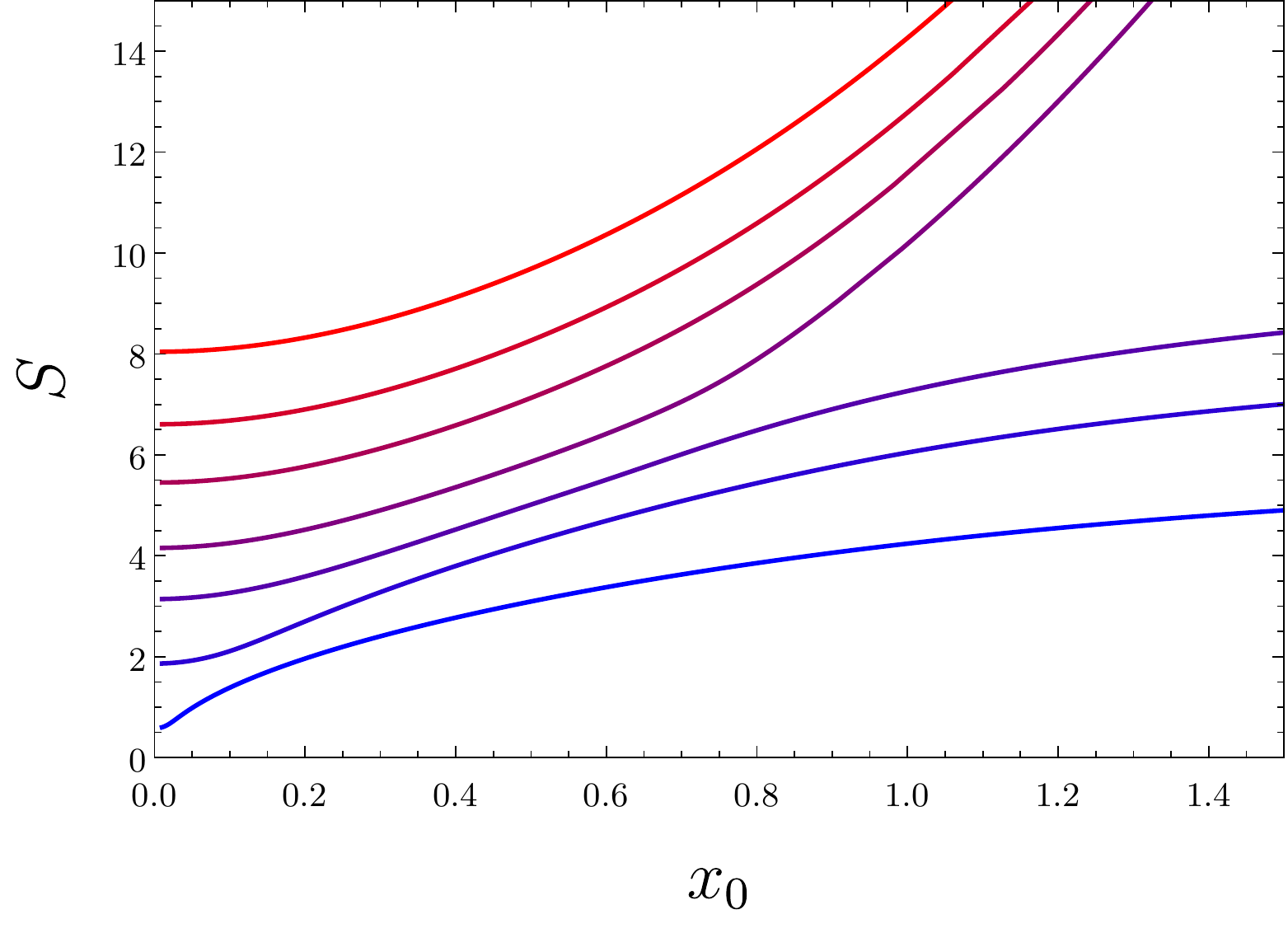}
\caption{Entropy of black holes that are smooth deformations of AdS$_2\times\mathbb{S}^2$ geometries as a function of $x_0$ for various values of $Q$. From blue to red we have $Q=0.43, 0.77, 1, 1.15, 3^{1/4}, 1.45, 1.6$. We work in units such that $\mu L^4=1$. For large enough $Q$, the curves tend to the Einstein gravity values in both limits $x_0\rightarrow 0,\infty$, but when $Q$ is too small the area tends to a constant value for $x_0\rightarrow\infty$.}
\label{fig:S}
\end{figure}

Now, using again Eqs.~\req{eq:constr1} we can study the entropy as a function of  $x_0$ and $Q$. For instance, in the limit $x_0<<Q$, we obtain the following approximate value,
\begin{equation}
S=\frac{\pi}{G}\left[Q^2+2x_0^2 \left(1+\frac{\mu  L^4}{Q^4}\right)+\frac{ 12 \mu L^4x_0^4}{Q^{10}} \left(\mu  L^4-2 Q^4\right)\right]\, ,
\end{equation}
while for large $x_0$ we have to distinguish between the two different possibilities,

\begin{equation}
S(x_0\rightarrow\infty)=
\begin{cases}
\frac{\pi}{G}\left(Q^2+2x_0^2\right)\hskip0.1cm&\text{if}\quad Q>Q_{\rm thr}\\
\frac{\pi \alpha}{G}\left(1+\frac{15\mu L^4}{\alpha^2(2\alpha+6Q^2)}\right)\hskip0.1cm&\text{if}\quad Q<Q_{\rm thr}\, ,
\end{cases}
\end{equation}
where $\alpha$ is the parameter that we introduced in \req{eq:x0large}. The complete profile of $S(x_0)$ for various values of the charge is shown in Fig.~\ref{fig:S}.

One disadvantage of this analysis is that, as we mentioned earlier, the parameter $x_0$ cannot be identified with the angular momentum, and therefore, the relation $S(x_0,Q)$ does not have a direct physical interpretation. Nevertheless, we can also study the entropy as a function of the area and of the charge, \textit{i.e}, $S(\mathcal{A},Q)$, and in this case the relation is meaningful since it involves three physically relevant quantities. In fact, it is interesting to check that the entropy is not only a function of the area, since it depends also on the relative amount of charge and angular momentum of the black hole.  Manipulating the equations in \req{eq:constr1}, we can write the entropy \req{eq:entro} in the following form,
\begin{equation}\label{entrolambda}
S=\frac{\mathcal{A}}{4G}\left[1+\frac{48\pi^2 \mu L^4 \lambda(\mathcal{A},Q)}{\mathcal{A}^2}\right]\, ,
\end{equation}
where $\lambda(\mathcal{A},Q)$ is a function obtained as a solution of the equation


\begin{equation}\label{lambdaeq}
\begin{aligned}
&12 \lambda ^3 \mu L^4  \left[\left(\frac{\mathcal{A}}{4\pi}\right)^2-\mu L^4 \right]+3 \lambda ^2 \mu L^4  \left[5 \mu L^4 -6 \left(\frac{\mathcal{A}}{4\pi}\right)^2\right]\\
&+\lambda  \left(\frac{\mathcal{A}}{4\pi}\right) \left[2 \left(\frac{\mathcal{A}}{4\pi}\right)^3+2 \left(\frac{\mathcal{A}}{4\pi}\right) \mu L^4 -3 \mu L^4  Q^2\right]\\
&+\left(\frac{\mathcal{A}}{4\pi}\right)^3 \left[Q^2-\left(\frac{\mathcal{A}}{4\pi}\right)\right]=0\, .
\end{aligned}
\end{equation}

\begin{figure}[t!]
	\includegraphics[width=\columnwidth]{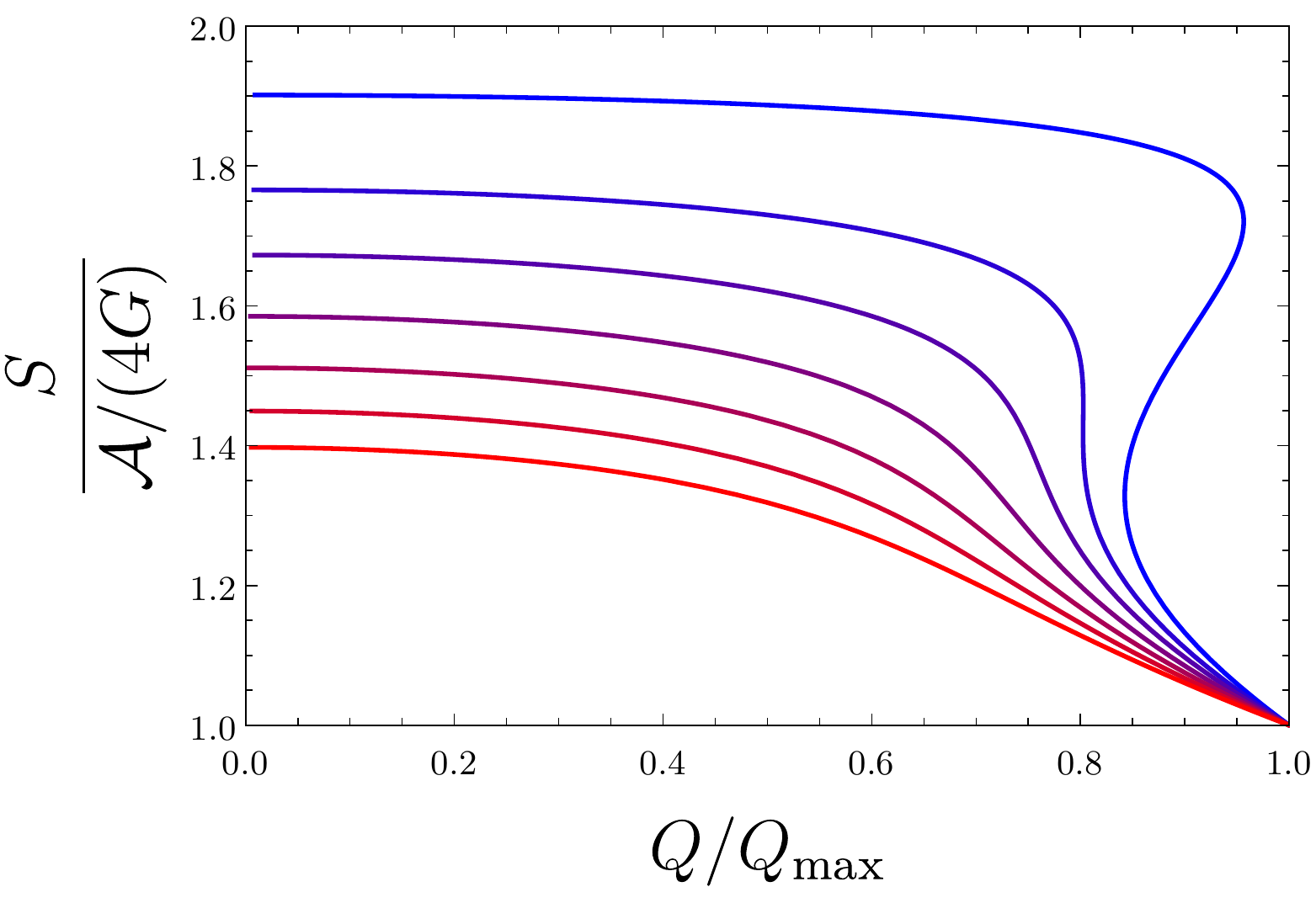}
	\caption{Entropy of black holes that are smooth deformations of AdS$_2\times\mathbb{S}^2$ geometries as a function of the charge for fixed values of the area. We plot the ratio $S/(\mathcal{A}/(4G))$ in order to facilitate the comparison between the different curves, while the charge is normalized by $Q_{\rm max}=\mathcal{A}/(4\pi)$. From blue to red we have $\mathcal{A}/(4\pi \sqrt{\mu}L^2)=1.8, 1.91, 2, 2.1, 2.2, 2.3, 2.4$. We observe the presence of a critical point where the curve starts being multivalued. }
	\label{fig:SAQ}
\end{figure}

On general grounds, for a fixed value of the area, the charge can vary from $Q=0$, which would correspond to a neutral rotating black hole, to $Q_{\rm max}^2=\mathcal{A}/(4\pi)$, in whose case there is no rotation and the solution is AdS$_2\times\mathbb{S}^2$.\footnote{When the area is sufficiently small we obtain solutions that have $Q>Q_{\rm max}$ --- see Fig.~\ref{fig:entrototal} --- but here we focus only in the case in which $Q$ ranges from $0$ to $Q_{\rm max}$ for simplicity.} It is then an interesting exercise to determine for which of these black holes of fixed area the entropy is maximal. In Fig.~\ref{fig:SAQ} we show the ratio $\frac{S}{\mathcal{A}/(4G)}$ as a function of the charge for several fixed values of the area. First, we observe that, indeed, the entropy does not only depend on the area, but also on the charge. For $Q=Q_{\rm max}$ we get $S=\mathcal{A}/(4G)$, since in that case the solution has no corrections. Nevertheless, when we decrease the charge leaving the area fixed --- which implies that we turn on the angular momentum --- the ratio between entropy and area increases. In all cases shown we see that, for a given area, a purely rotating black hole is the one that stores more information. 
We also observe an interesting phenomenon taking place when the area is small enough: if $\mathcal{A}<1.91\times 4\pi\sqrt{\mu} L^2$ the corresponding curve becomes multivalued, indicating the existence of several black holes with same area and charge, but different entropy. This suggests the presence of a phase transition from the black hole of smaller entropy to the one of larger entropy. In that case, we see that the phase space would contain a critical point at $\mathcal{A}_{\rm cr}\approx 1.91\times 4\pi\sqrt{\mu} L^2$, $Q_{\rm cr}\approx 1.11 \mu^{1/4} L$, $S_{\rm cr}\approx 8.63 \frac{\sqrt{\mu}L^2}{G}$. This picture is not completely accurate, though, because one should fix the angular momentum instead of the area in order to compare different solutions, and also because at zero temperature one cannot speak of phase transitions. Nevertheless, this result does suggest that some sort of decay could take place from one type of solution to another.

\section{Additional solutions}\label{sec:add}
\begin{figure}[ht]
\includegraphics[scale=0.5]{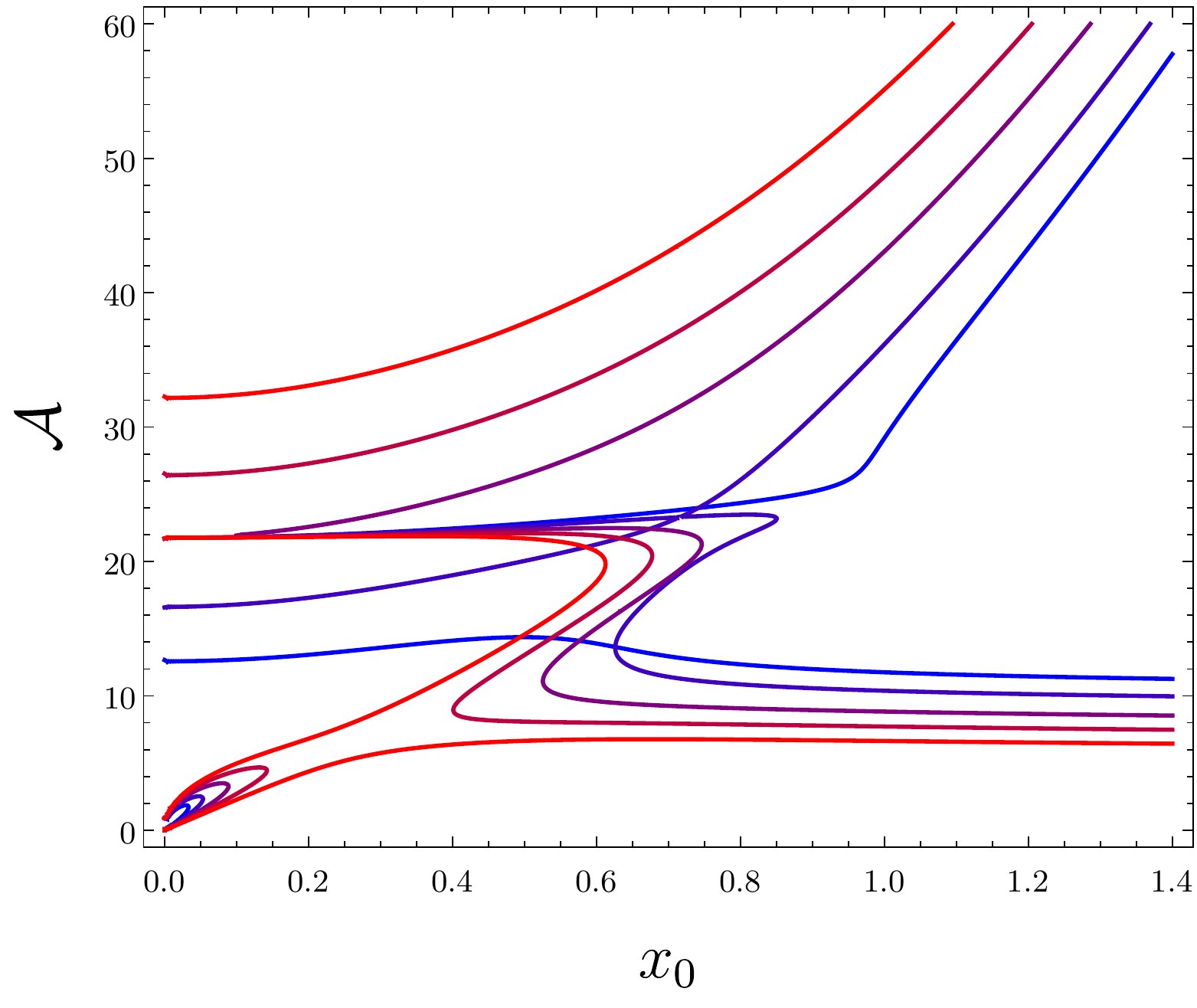}
\caption{Black hole area as a function of $x_0$ for various values of $Q$. We include all the branches of solutions. From blue to red we have $Q=1, 1.15, 3^{1/4}, 1.45, 1.6$. We work in units such that $\mu L^4=1$.}
\label{fig:areaall}
\end{figure}

In the previous section we focused on the branch of solutions that are smoothly connected to an AdS$_2\times\mathbb{S}^2$ geometry, since these are particularly relevant --- and the only ones that exist in Einstein gravity. However, when we solve the system of equations \req{eq:constr1} we observe that other solutions for $n(x_0,Q)$ and $\omega(x_0, Q)$ exist. A useful way of visualizing the space of solutions is to study the relation $\mathcal{A}(x_0)$ for fixed values of the charge, which we show in Fig.~\ref{fig:areaall}. This plot contains the curves that we showed in Fig.~\ref{fig:area}, but we see that new branches appear. In fact, for fixed values of $Q$ and $x_0$ there can be up to four different solutions, which represent black holes with very different properties. 

\subsection{Branches of solutions}
In the limit of $x_0\rightarrow\infty$, we observe that there are only two possible solutions; one which recovers the properties of extremal Kerr-Newman black holes --- in particular, $\mathcal{A}\rightarrow4\pi(Q^2+2x_0^2)$ --- and another one whose area tends to a constant --- see Eq. \req{eq:x0large}. On the other hand, near $x_0=0$ we have in general four different solutions, which can be obtained by assuming different expansions of the parameters $n$ and $\omega$, as we show in the Appendix \ref{app:termo}.  One of them belongs to the AdS$_2\times\mathbb{S}^2$ branch that we studied in the previous section, so we will now analyze the additional solutions.

\allowdisplaybreaks
\subsubsection{Branch A}
One possible solution of the equations \req{eq:constr1} yields
\begin{align}
n^2=&\frac{x_0^4 \left(2 \sqrt{3\mu } L^2+3 Q^2\right)}{18 \mu  L^4}+\mathcal{O}(x_0^6)\,\\
\mathcal{A}=&4 \pi \left[\sqrt{3\mu } L^2+ x_0^2 \left(\frac{1}{2}-\frac{Q^2}{4\sqrt{3\mu } L^2}\right)+\mathcal{O}(x_0^4)\right]\,
\end{align}
where we recall that $\mathcal{A}=4\pi x_0/\omega$. It is important to note that the near-horizon geometry corresponding to this choice of parameters exists for arbitrarily small values of $x_0$, but not for $x_0=0$. One remarkable fact about this solution is that in the limit of $x_0\rightarrow 0$ the area tends to a constant value which is independent of the charge. On the other hand, the entropy can be computed using \req{eq:entro}, and we obtain
\begin{align}
S=&\frac{2\pi}{G}\left[ \sqrt{3\mu } L^2- x_0^2 \left(\frac{1}{6}+\frac{Q^2}{4\sqrt{3\mu } L^2}\right)+\mathcal{O}(x_0^4)\right]
\end{align}
Thus, the entropy also tends to a universal constant value in the limit of vanishing $x_0$, which interestingly enough corresponds to $\mathcal{A}/(2G)$.  Observe that, for fixed values of $Q$ and $x_0$, this solution can be entropically favoured with respect to the one belonging to the AdS$_2\times\mathbb{S}2$ branch. In fact, we get the following condition for small values of $x_0$:
\begin{equation}
S_A>S_{\text{AdS$_2\times\mathbb{S}_2$}} \Leftrightarrow Q^2<2\sqrt{3\mu}L^2-\frac{7x_0^2}{2}+\ldots.
\end{equation}
However, this is not enough in order to argue that a transition from one solution to another will take place when that bound is saturated, since the angular momentum could depend differently on $x_0$ in both solutions, and hence, we would be comparing black holes with different conserved charges. In fact, this solution has $x_0/n\sim 1/x_0$ when $x_0\rightarrow 0$, which implies that the geometry departs largely from AdS$_2\times\mathbb{S}^2$, and this suggests that it actually could have a large angular momentum.  

Finally, let us comment on how the black holes in this branch behave as we increase the angular momentum. Looking at Fig.~\ref{fig:areaall} we observe three possibilities. If the charge is large enough, there is a maximum value of $x_0$ for which we can extend the branch, and at this point it merges with branch C. If the charge is smaller, the branch is connected to the solutions that have a finite area in the limit $x_0\rightarrow\infty$, and if it is small enough ($Q<Q_{\rm thr}\approx 1.13\mu^{1/4} L$), it is connected to the Kerr-Newman branch. In other words, this implies that if we take an initial black hole with little charge but large area and angular momentum, the black hole will approach one of the solutions in this branch as it losses angular momentum, instead of an AdS$_2\times\mathbb{S}^2$ geometry.  
  
\begin{figure}[t!]
\includegraphics[width=\columnwidth]{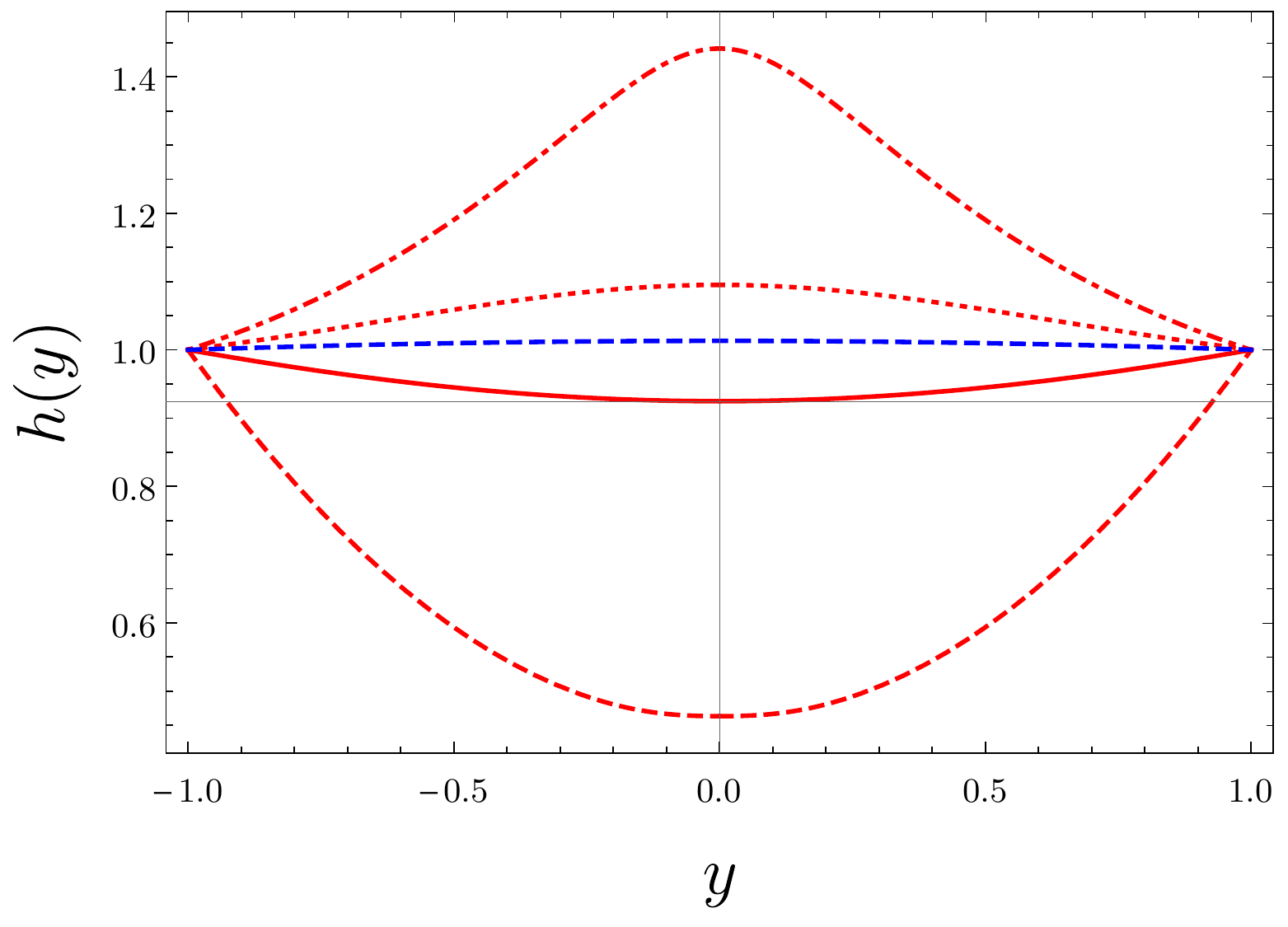}
\caption{Different near-horizon geometries with $Q=(9\mu)^{1/4} L$ and $x_0=0.2 \mu^{1/4} L$. In each case, we show the quantity $h(y)\equiv\frac{x_0g(y)}{\omega(1-y^2)}$, which allows for a simpler comparison between the several curves. Solid red line: AdS branch. 
Red dashed line: branch A. 
 Red dot-dashed line: branch B. 
Red dotted line: branch C. 
 Blue dashed line: Kerr-Newman case. }
\label{fig:othersol}
\end{figure}
  
\subsubsection{Branch B}
The second additional solution has the following values of $n^2$ and $\mathcal{A}$
\begin{align}
n^{2}=&x_{0} \frac{\sqrt{3\mu } L^2}{Q}+x_{0}^2 \left(\frac{25 \mu  L^4}{6 Q^4}-\frac{4}{5}\right)+\mathcal{O}(x_0^3)\,\\
\mathcal{A}=&4\pi\left[  x_{0}\frac{5 \sqrt{\mu/3 } L^2 }{ Q}+x_{0}^2\ \left(\frac{25 \mu  L^4}{2Q^4}-\frac{5}{3}\right)+\mathcal{O}(x_0^3)\right]\, .
\end{align}
In this case, the area tends to zero independently of the charge when $x_0\rightarrow 0$. Also, unlike in the previous case, we have $x_0/n\rightarrow 0$, which we can interpret as a sign that the geometry is indeed slowly rotating. Now, the most interesting fact about this branch of solutions is that,  even though the area vanishes in the limit of $x_0\rightarrow 0$, their entropy remains finite, namely

\begin{equation}
S=\frac{3 \pi  Q^2}{5 G}+x_{0}\ 2\pi\frac{ 6 Q^4-25 \mu  L^4}{25 G \sqrt{3\mu } L^2 Q}+\mathcal{O}(x_0^2)\, .
\end{equation}
Thus, the entropy per unit area in these black holes becomes arbitrarily large. 

\subsubsection{Branch C}
The third and last additional solution allows for a series expansion in powers of $x_0^{1/2}$, and the leading terms for $n^2$, area and entropy read 
\begin{align}
n^{2}=&x_{0}\frac{\sqrt{6\mu} L^2}{Q}-\left(\sqrt{x_{0}}\right)^{3}\frac{\left(2/3\right)^{1/4} \mu ^{3/4} L^3}{Q^{5/2}}\, ,\\
\mathcal{A}=&\sqrt{x_{0}} \ 2^{3/4} {3}^{1/4} 2 \pi  {\mu }^{1/4} L \sqrt{Q} -x_{0}\frac{ \pi  \sqrt{6\mu } L^2}{Q}\, ,\\
S=&\sqrt{x_{0}} \ \frac{2^{3/4} {3}^{1/4} \pi \ {\mu }^{1/4} L \sqrt{Q}}{G}+x_{0} \ \frac{\pi  \sqrt{\mu/6 } L^2 }{ G Q}\, ,\\
\end{align}
Note that, again, $x_0/n\rightarrow 0$, so that this solution can actually be slowly rotating, while the entropy tends to $S\rightarrow \mathcal{A}/(2G)$.

Once the desired branch is chosen, it is possible to solve the equation \req{mastereq} numerically in order to obtain the profile of $g(y)$, as we explained previously. A comparison between these solutions is shown in Fig.~\ref{fig:othersol}. 

\subsection{Entropy as a function of area and charge}
\begin{figure}[ht]
	\includegraphics[width=\columnwidth]{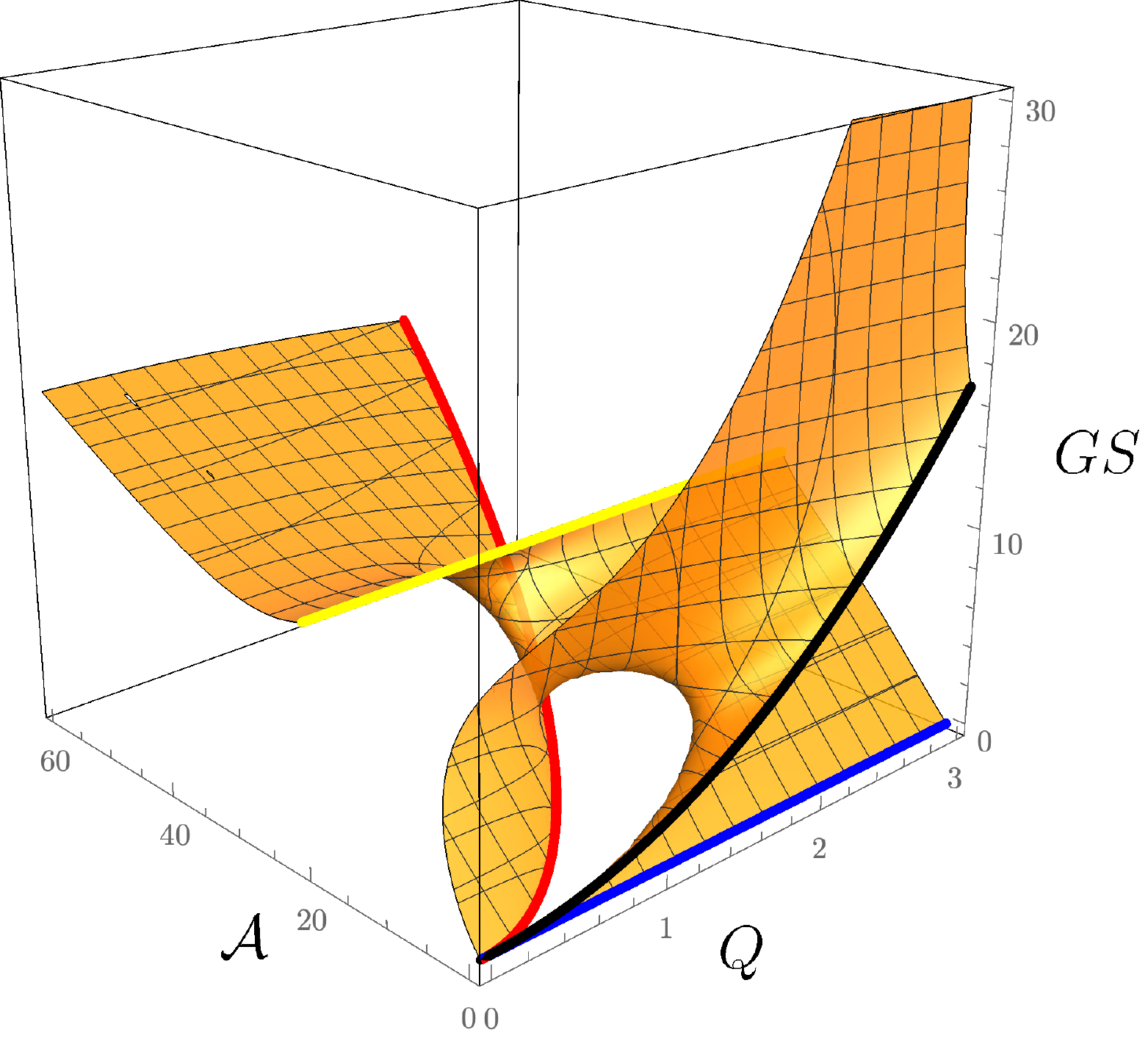}
	\caption{Black hole entropy as a function of the area and charge. The thick color lines represent the different $x_0\rightarrow 0$ limits: the red line corresponds to the AdS$_2\times \mathbb{S}^2$ solutions, while yellow, black and blue lines correspond to branches  A, B and C, respectively. We work in units such that $\mu L^4=1$. }
	\label{fig:entrototal}
\end{figure}

The preceding analysis is useful in order to characterize the space of near-horizon geometries of ECG, but it has the disadvantage that we cannot interpret $x_0$ as the spin parameter $a$. Thus, it is more meaningful to study the relation $S(\mathcal{A},Q)$, which we can find exactly by using Eqs.~\req{entrolambda} and \req{lambdaeq}. In Fig.~\ref{fig:SAQ} we only plotted part of this relation. The complete structure of $S(\mathcal{A},Q)$ including all the solutions is quite involved and we show it as a 3-dimensional plot in Fig.~\ref{fig:entrototal}. 
In obtaining this surface we have taken into account that the solutions of Eq.~\req{lambdaeq} must be such that $n^2>0$ and $x_0^2>0$. The red line corresponds to the AdS$_2\times \mathbb{S}^2$ geometries, and interestingly these are the only ones for which $S=\mathcal{A}/(4G)$ --- any other solution has $S>\mathcal{A}/(4G)$.  We also represent the various $x_0\rightarrow 0$ limits, which correspond the yellow, black and blue curves (for branches A, B and C respectively).

As we can see, for large enough horizon area, the surface in Fig.~\ref{fig:entrototal} has only one branch, which recovers the Einstein gravity behaviour when $\mathcal{A}\rightarrow\infty$. Now, imagine that we take one of these large black holes and we start decreasing the area leaving the charge fixed --- this could be interpreted as the black hole radiating away the angular momentum.\footnote{This picture is not completely accurate because we are moving in the space of extremal black holes. Thus, one should imagine that energy is emitted along with angular momentum, so that we keep the black hole extremal, or near-extremal, during the process.} We find that there are two possible endpoints of this process: if the charge is large enough, then at some point we hit an AdS$_2\times\mathbb{S}^2$ geometry and the black hole has radiated all the angular momentum. In order to continue evaporating it must now lose charge. On the other hand, if the charge is too small (as we saw earlier, $Q<Q_{\rm thr}\approx 1.13\mu^{1/4} L$), we approach the yellow line, which corresponds to the $x_0\rightarrow 0$ limit of branch A, and for which $\mathcal{A}=4\pi\sqrt{3\mu} L^2$. Interestingly, in this situation the area and the entropy of the black hole remain constant even if it loses (or gains) charge. Thus, the final product of black hole evaporation is quite different depending on which path we follow in the phase space. We also observe that for small $\mathcal{A}$ the surface $S(\mathcal{A},Q)$ is multivalued, hence transitions or decays between solutions might occur.  This illustrates that the phase space of (extremal) black hole solutions may become quite complicated in higher-derivative gravity.

\section{Discussion}\label{sec:disc}
In this paper we have provided the first non-perturbative examples of near-horizon geometries of rotating black holes in higher-order gravity. This has been possible thanks to the special form of the equations of motion of Einsteinian cubic gravity --- the density given by \req{eq:P} --- which can be reduced to a single second-order differential equation for one variable. Even more striking, we have been able to obtain the area and the entropy exactly in terms of the parameters of the solution, and in particular, we found the relation between black hole area, charge and entropy, $S(\mathcal{A},Q)$ --- see Eqs. \req{entrolambda} and \req{lambdaeq}.
It must be noted that obtaining these quantities analytically is not possible in general higher-order theories, where the simplification of the equations that we reported does not take place. However, we do expect that there is a subset of Generalized Quasi-topological theories for which the same simplification takes place. This subset will correspond to the same type of theories admitting taub-NUT solutions that was studied in \cite{Bueno:2018uoy}, where, in particular, a quartic four-dimensional density of this kind was constructed. We expect that higher-order versions of these densities exist as well, and it would be interesting to study extremal near-horizon geometries in this family of theories, thus generalizing the results presented here. In fact, we believe that the higher-order generalizations could solve some of the difficulties that we have found in our analysis and that we discuss next. 

\subsection{Large angular momentum?}
Perhaps the most worrisome problem we have found is that the equation \req{mastereq} seems to have no smooth solutions when the angular momentum is large compared to the charge. In particular, purely rotating (regular) black holes do not exist even in the regime where the corrections are supposed to be small. The reason, as we explained, is the vanishing of the coefficient of $g''$ in Eq.~\req{mastereq} at some point, which implies that the solution will not be smooth there. This issue could go away for higher-order densities, or for some appropriate combination of those, and it would be interesting to explore this possibility. On the other hand, this problem could be related to the fact that we are dealing with extremal black holes. It is known that there are certain difficulties associated with extremality (\textit{e.g.}, the instability of horizons \cite{Aretakis:2012ei}), and these arise explicitly in the case of higher-derivative theories --- we further comment on this below. Therefore, it might happen that the problem of non-existence only affects to extremal black holes, but that (arbitrarily) near-extremal ones are fine.
Despite this drawback, we believe the values found for the entropy and area of these black holes are meaningful even in the region of parameter space where no solution seems to exist. Indeed, from the point of view of EFT one should assume a perturbative expansion of the solution, and in this scheme the issue in the differential equation \req{mastereq} disappears.  Thus, at least the perturbative corrections to the entropy,
\begin{equation}
S=\frac{\mathcal{A}}{4G}\left[1+\frac{24\pi^2 \mu L^4  (\mathcal{A}-4 \pi Q^2) }{\mathcal{A}^3}+\mathcal{O}(L^8)\right]\, ,
\end{equation}
should be meaningful in the full parameter space.  


\subsection{Multiplicity of solutions}
Paradoxically enough,  when Eq.~\req{mastereq} allows for solutions, it has many. We have seen that for fixed values of $x_0$ and $Q$, we have usually several branches of solutions with different values of the area and the entropy. But we also observed that, even when the corresponding branch has been chosen, the equation \req{mastereq} can have several solutions. This is, we can have different near-horizon geometries with the same values of the charge, $x_0$, area and entropy, which only differ in the shape of the horizon. Thus, in Sec. \ref{sec:AdS2} we only constructed numerically the solutions that are smooth deformations of AdS$_2\times\mathbb{S}^2$ geometries, but in general there are more solutions which are characterized by the same set of integration constants. In particular, the equation \req{eq:g0eq} corresponding to the limit $x_0\rightarrow 0$ seems to have an increasing number of solutions as $Q$ grows. This means that there are solutions of the form AdS$_2\times\mathcal{M}_2$, where $\mathcal{M}_2$ is not a sphere, but nonetheless all of these solutions have the same area and entropy. A similar situation occurs for finite $x_0$. While this is an interesting phenomenon, a thorough classification of these solutions would considerable enlarge the present manuscript, and thus these additional solutions could be studied elsewhere. 
The degeneracy of solutions seems to be related to the sign of the higher-order coupling $\mu$, and it would have not appeared had we taken $\mu<0$. The reason for taking $\mu>0$ is that this is required in order for asymptotically flat/AdS black holes to exist \cite{PabloPablo2}. However, it is possible that for other higher-order densities the sign that allows for black holes is the same that would yield unicity of near-horizon geometries.\footnote{A similar phenomenon has been observed in the cosmological context, where the appropriate sign for black holes in ECG is the opposite to the one required in order to produce inflation. However, for quartic densities (and in general, densities containing even powers of the curvature), both signs agree \cite{Arciniega:2018tnn}.}

\subsection{Global solutions?}
Another relevant question is whether there exist global black hole solutions (containing an asymptotic region) of which the solutions we have constructed are the near-horizon limit. Although it may appear shocking at first, we do not expect those solutions to exist. The reason is that the boundary problem in higher-derivative gravity is not well-posed in the presence of a degenerate horizon. This is more easily understood in the case of static, charged black holes, which allow for a simple description in ECG. Those solutions were briefly discussed in \cite{PabloPablo2}, where, similarly to the case here, it was shown that the equations of motion reduce to a second-order equation for one variable. Then, one has to impose a boundary condition at infinity and another one at the horizon, and this fixes the solution. But when the horizon is degenerate, the condition at the horizon turns out to fix two integration constants and it is not possible to demand the asymptotic condition. Hence, no black hole solutions exist in that case. Nevertheless, arbitrarily near-extremal ones exist, and we expect that the same behaviour will be found in the rotating case. Hence, the near-horizon geometries we have constructed make sense as a limit that non-extremal black holes can approach, but never reach. In particular, the area and entropy (and also the shape) of non-extremal black holes will tend to those found here when they approach extremality.

\subsection{Asymptotic charges}
Finally, one limitation of the near-horizon analysis is that we lose the information about the mass and the angular momentum of these black holes. We argued that the variables $x_0$ and $n$ would be related, respectively, to the spin $a$ and to the mass $M$ but we lack a precise relation. Knowing the values of $a$ and $M$ would be very interesting in order to study  corrections to the extremality bound and to determine the relation between the entropy and the physical charges, $S(a,Q)$. A possible direction to achieve this goal would entail finding a generalization of Komar charge for higher-order gravities that would allow us to write the asymptotic charges as an integral over the horizon \cite{Kastor:2008xb}.

\acknowledgments
We are pleased to thank Pablo Bueno, Roberto Emparan and Robie Hennigar for useful comments.  PAC was mostly funded by Fundaci\'on la Caixa through a ``la Caixa - Severo Ochoa'' International pre-doctoral grant. In the last stages of this project PAC was supported by the KU Leuven grant ``Bijzonder Onderzoeksfonds C16/16/005 – Horizons in hoge-energie fysica".
DP is funded by a ``Centro de Excelencia Internacional UAM/CSIC'' FPI pre-doctoral grant. PAC and DP were further supported by the MINECO/FEDER, UE grant PGC2018-095205-B-I00, and by the ``Centro de Excelencia Severo Ochoa" Program grant  SEV-2016-0597.

\appendix
\allowdisplaybreaks
\onecolumngrid
\section{Equations of motion}\label{app:eoms}
When evaluated on $N(x)=1$,  the gravitational tensor in \req{eq:fe} has the following  non-vanishing components:
\begin{eqnarray}
\mathcal{E}_{\psi\psi}&=&f(x)^2\mathcal{E}_{xx}\, ,\\
\mathcal{E}_{\psi t}&=&-2nr f(x)^2\mathcal{E}_{xx}\, ,\\
\mathcal{E}_{tt}&=&-r^4\mathcal{E}_{rr} +4n^2r^2 f(x)^2\mathcal{E}_{xx}\, .
\end{eqnarray}
In addition, we can relate $\mathcal{E}_{rr}$ to $\mathcal{E}_{xx}$ thanks to the Bianchi identity $\nabla^{\mu}\mathcal{E}_{\mu\nu}=0$,
\begin{equation}
\mathcal{E}_{rr}=\frac{\left(n^2+x^2\right)}{2x r^2} \left(f(x) \left(n^2+x^2\right) \frac{d\mathcal{E}_{xx}}{dx}+\mathcal{E}_{xx}\left(\left(n^2+x^2\right) f'(x)+2 x f(x)\right)\right)\, .
\end{equation}
Thus, everything is determined by the component $\mathcal{E}_{xx}$, which reads

\begin{align}
f\mathcal{E}_{xx}=&\Lambda+\frac{f \left(-n^2+x^2\right)+\left(n^2+x^2\right)(1+x f')}{\left(n^2+x^2\right)^2} +L^4 \mu  \Bigg[-\frac{3 f^3 \left(n^6+16 n^4
	x^2-45 n^2 x^4\right)}{\left(n^2+x^2\right)^6}+\left(\frac{3 f x^3}{\left(n^2+x^2\right)^4}\right.\\&\left.+\frac{3 f^2 \left(3 n^4 x-62 n^2
	x^3+x^5\right)}{\left(n^2+x^2\right)^5}\right) f'+\left(\frac{3 \left(n^2-x^2\right)}{4 \left(n^2+x^2\right)^3}+\frac{3 f \left(n^4+37 n^2 x^2-2
	x^4\right)}{2 \left(n^2+x^2\right)^4}\right) f'^2-\frac{3 n^2 x f'^3}{2 \left(n^2+x^2\right)^3}\\&+\left(-\frac{3 f
	\left(n^2+2 x^2\right)}{2 \left(n^2+x^2\right)^3}+\frac{f^2 \left(6 n^4+45 n^2 x^2-3 x^4\right)}{\left(n^2+x^2\right)^4}+\frac{3 f \left(-5 n^2
	x+x^3\right) f'}{\left(n^2+x^2\right)^3}\right) f''-\frac{3 f x^2 \left(f''\right)^2}{4 \left(n^2+x^2\right)^2}\\&+\left(\frac{3 f^2 x \left(-4
	n^2+x^2\right)}{2 \left(n^2+x^2\right)^3}+\frac{3 f x}{2 \left(n^2+x^2\right)^2}-\frac{3 f x^2 f'}{4 \left(n^2+x^2\right)^2}\right) f^{(3)}\Bigg]\, .
\end{align}
The electromagnetic energy-momentum tensor has the same structure and hence the equations of motion are reduced to $\mathcal{E}_{xx}=T_{xx}$.

\onecolumngrid

\section{Solution of the Thermodynamic Quantities}\label{app:termo}
Three of the four branches of solutions of the constraint equations \eqref{eq:constr1} belong to the following class,
\begin{equation}\label{class1}
n(x_{0},Q)=x_{0}^{\alpha}\sum_{k=0}^{\infty}n_{k}(Q)x_{0}^{k}\, ,\quad \omega(x_{0},Q)=x_{0}^{\beta}\sum_{k=0}^{\infty}\omega_{k}(Q)x_{0}^{k},
\end{equation}
where $(\alpha,\beta)$ are real parameters and we assume $n_{0}(Q)\ne0$ and $\omega_{0}(Q)\ne 0$. The choice $(\alpha,\beta)=(0,1)$ corresponds to the AdS$_{2}\times \mathbb{S}^{2}$ branch, while the choices $(\alpha,\beta)=(2,1)$ and $(\alpha,\beta)=(1/2,0)$ lead to other two solutions. The remaining branch belongs to the class
\begin{equation}\label{class2}
n(x_{0},Q)=\sum_{k=0}^{\infty}n_{k}(Q)\left(\sqrt{x_{0}}\right)^{k}\, ,\quad \omega(x_{0},Q)=\sum_{k=0}^{\infty}\omega_{k}(Q)\left(\sqrt{x_{0}}\right)^{k}.
\end{equation}
The solution for all coefficients in each of the expansions can be found explicitly. In the following we exhibit the first four terms of the solutions for $n^2$ and $\omega$, as well as four terms of the corresponding expansions of the area $\mathcal{A}$, Wald entropy $S$, and relative entropy $S/S_{\rm BH}$, where $S_{\rm BH}=\mathcal{A}/(4 G)$.

For the branch corresponding to AdS$_{2}\times\mathbb{S}^{2}$, determined by the coefficients $(\alpha,\beta)=(0,1)$, 
\begin{align}
n^{2}=&Q^{2}+x_{0}^{2} \left(\frac{\mu  L^4}{Q^4}+1\right)+x_{0}^{4} \ \frac{2 \mu  L^4  \left(11 \mu  L^4-13 Q^4\right)}{Q^{10}}+x_{0}^{6} \ \frac{3 \mu  L^4  \left(25 \mu ^2 L^8-90 \mu  L^4 Q^4+56 Q^8\right)}{Q^{16}}\\
\omega=&x_{0}\frac{1}{Q^2}-x_{0}^{3}\ \frac{ 4 Q^4-2 \mu  L^4}{2 Q^8}+x_{0}^{5}\ \frac{ -11 \mu ^2 L^8+8 \mu  L^4 Q^4+4 Q^8}{Q^{14}}-x_{0}^{7}\ \frac{4 \left(26 \mu ^3 L^{12}-69 \mu ^2 L^8 Q^4+36 \mu  L^4 Q^8+2 Q^{12}\right)}{Q^{20}}\\
\mathcal{A}=&4 \pi  Q^2+\pi  x_{0}^{2} \left(8-\frac{4 \mu  L^4}{Q^4}\right)+x_{0}^{4}\ \frac{48 \pi  \mu  L^4 \left(\mu  L^4-Q^4\right)}{Q^{10}}+x_{0}^{6}\ \frac{12 \pi  \mu  L^4  \left(27 \mu ^2 L^8-70 \mu  L^4 Q^4+36 Q^8\right)}{Q^{16}}\\
S=&\frac{\pi  Q^2}{G}+x_{0}^{2} \ \frac{2 \pi \left(\mu  L^4+Q^4\right)}{G Q^4}+x_{0}^{4}\ \frac{12 \pi  \mu  L^4 \left(\mu  L^4-2 Q^4\right)}{G Q^{10}}-x_{0}^{6}\ \frac{6 \pi  \mu  L^4 \left(3 \mu ^2 L^8+16 \mu  L^4 Q^4-24 Q^8\right)}{G Q^{16}}\\
S/S_{\rm BH}=&1+x_{0}^{2}\ \frac{3 \mu  L^4 }{Q^6}+x_{0}^{4}\ \frac{3 \mu  L^4  \left(\mu  L^4-6 Q^4\right)}{Q^{12}}+x_{0}^{6}\ \frac{6 \mu  L^4  \left(-22 \mu ^2 L^8+21 \mu  L^4 Q^4+12 Q^8\right)}{Q^{18}}\,.
\end{align}

For the branch determined by the coefficients $(\alpha,\beta)=(2,1)$,
\begin{align}
n^{2}=&x_{0}^{4}\ \frac{ 2  \sqrt{3\mu } L^2+3 Q^2}{18 \mu  L^4}+x_{0}^{6} \ \frac{ 2 \mu  L^4- \sqrt{3\mu } L^2 Q^2+9 Q^4}{108 \mu ^2 L^8}\\ \notag
&+x_{0}^{8}\ \frac{ 328 \mu ^2 L^8+612 \sqrt{3} \mu ^{3/2} L^6 Q^2+330 \mu  L^4 Q^4+117  \sqrt{3\mu } L^2 Q^6+432 Q^8}{5184 \sqrt{3} \mu ^{7/2} L^{14}+7776 \mu ^3 L^{12} Q^2}\\ 
\omega=&x_{0}\ \frac{1}{\sqrt{3\mu } L^2}+x_{0}^{3}\ \frac{ \sqrt{3} Q^2-6 \sqrt{\mu}L^2}{36 \mu^{3/2}L^6}-x_{0}^{5}\ \frac{ -100 \sqrt{3} \mu  L^4+72 \sqrt{\mu } L^2 Q^2-33 \sqrt{3} Q^4}{2592 \mu ^{5/2} L^{10}}\\ \notag
&-x_{0}^{7}\ \frac{ 3584 \sqrt{3} \mu ^2 L^8+4344 \mu ^{3/2} L^6 Q^2+636 \sqrt{3} \mu  L^4 Q^4+198 \sqrt{\mu } L^2 Q^6-765 \sqrt{3} Q^8}{31104 \mu ^{7/2} L^{14} \left(2  \sqrt{3\mu } L^2+3 Q^2\right)}\\
\mathcal{A}=& 4 \pi  \sqrt{3\mu } L^2 +x_{0}^{2}\ \pi \left(2-\frac{Q^2}{\sqrt{3\mu } L^2}\right)- x_{0}^{4}\ \pi \frac{  28 \mu  L^4+27 Q^4}{72 \sqrt{3} \mu ^{3/2} L^6}\\ \notag
&+ x_{0}^{6}\ \pi \frac{ 2048 \mu ^2 L^8+1224 \sqrt{3} \mu ^{3/2} L^6 Q^2+588 \mu  L^4 Q^4-246 \sqrt{3\mu } L^2 Q^6-585 Q^8}{2592 \left(2 \mu ^3 L^{12}+\sqrt{3} \mu ^{5/2} L^{10} Q^2\right)}\\
S=&\frac{2 \pi  \sqrt{3\mu } L^2}{G} -x_{0}^{2}\ \frac{2 \pi  L^2+\pi  \sqrt{3/\mu} Q^2}{6 G L^2}+ x_{0}^{4}\ \frac{\pi  \left(68 \sqrt{3} \mu  L^4+48 \sqrt{\mu } L^2 Q^2-27 \sqrt{3} Q^4\right)}{432 G \mu ^{3/2} L^6}\\ \notag
&- x_{0}^{6}\ \frac{\pi   \left(640 \sqrt{3} \mu ^2 L^8+2104 \mu ^{3/2} L^6 Q^2+284 \sqrt{3} \mu  L^4 Q^4-42 \sqrt{\mu } L^2 Q^6+195 \sqrt{3} Q^8\right)}{1728 G \mu ^{5/2} L^{10} \left(2 \sqrt{3\mu } L^2+3 Q^2\right)}\\
S/S_{\rm BH}=&2-x_{0}^{2}\ \frac{4 }{3  \sqrt{3\mu } L^2}+x_{0}^{4}\ \frac{4}{9 \mu  L^4}-x_{0}^{6}\ \frac{ 280 \mu ^{3/2} L^6+180 \sqrt{3} \mu  L^4 Q^2+54 \sqrt{\mu } L^2 Q^4-3 \sqrt{3} Q^6}{216 \left(2 \sqrt{3} \mu ^3 L^{12}+3 \mu ^{5/2} L^{10} Q^2\right)}
\end{align}

For the branch determined by the coefficients $(\alpha,\beta)=(1/2,0)$
\begin{align}
n^{2}=&x_{0} \frac{\sqrt{3\mu } L^2}{Q}+x_{0}^2 \left(\frac{25 \mu  L^4}{6 Q^4}-\frac{4}{5}\right)+x_{0}^{3}\frac{ 128125 \mu ^2 L^8-32700 \mu  L^4 Q^4+324 Q^8}{1800 \sqrt{3\mu } L^2 Q^{7}}\\ \notag
&+x_{0}^{4} \left(\frac{53125 \mu ^2 L^8}{81 Q^{10}}-\frac{4745 \mu  L^4}{27 Q^6}+\frac{16 Q^2}{375 \mu  L^4}+\frac{28}{3 Q^2}\right)\\
\omega=&\frac{\sqrt{3/\mu} Q}{5 L^2}-x_{0}\left(\frac{3}{2Q^2}-\frac{ Q^2}{5 \mu  L^4}\right)+x_{0}^{2}\ \frac{-216875 \mu ^2 L^8+33300 \mu  L^4 Q^4+1476 Q^8 }{9000 \sqrt{3} \mu ^{3/2} L^6 Q^{5}}\\ \notag
&- x_{0}^{3}\left(-\frac{64 Q^4}{1875 \mu ^2 L^8}+\frac{2}{15 \mu  L^4}+\frac{12475 \mu  L^4}{54 Q^8}-\frac{1141}{27 Q^4}\right)\\
\mathcal{A}=&x_{0}\ \pi\frac{20   \sqrt{\mu/3 } L^2 }{ Q}+x_{0}^2\ \pi \left(\frac{50 \mu  L^4}{Q^4}-\frac{20}{3}\right)+x_{0}^{3}\ \pi\frac{318125 \mu ^2 L^8-60300 \mu  L^4 Q^4+324 Q^8}{270 \sqrt{3\mu } L^2 Q^{7}}
\\ \notag
&+x_{0}^{4}\ 8\pi \frac{ 3203125 \mu ^2 L^8+72 \left(Q^{12}/\mu L^4\right)-743125 \mu  L^4 Q^4+27675 Q^8}{2025 Q^{10}}\\
S=&\frac{3 \pi  Q^2}{5 G}+x_{0}\ 2\pi\frac{ 6 Q^4-25 \mu  L^4}{25 G \sqrt{3\mu } L^2 Q}+x_{0}^{2}\ \pi\frac{ -18125 \left(\mu  L^4/Q^4\right)+108\left(Q^4/\mu  L^4\right)+6150}{1125 G}\\ \notag
&+x_{0}^{3}\ \pi\frac{ \left(-34203125 \mu ^3 L^{12}+6656250 \mu ^2 L^8 Q^4-283500 \mu  L^4 Q^8+1944 Q^{12}\right)}{67500 \sqrt{3} G \mu ^{3/2} L^6 Q^{7}}\\
S/S_{\rm BH}=&\frac{1}{x_{0}}\frac{3 \sqrt{3/\mu} Q^{3}}{25 L^2}+\frac{27 Q^4}{125 \mu  L^4}-\frac{13}{10}+x_{0}\frac{ -21125 \mu ^2 L^8+4380 \mu  L^4 Q^4+252 Q^8}{1000 \sqrt{3} \mu ^{3/2} L^6 Q^{3}}\\ \notag
&+x_{0}^{2} \left(\frac{224 Q^6}{3125 \mu ^2 L^8}-\frac{3595 \mu  L^4}{18 Q^6}+\frac{64 Q^2}{125 \mu  L^4}+\frac{82}{3 Q^2}\right)
\end{align}
Finally, for the class of solutions defined by \eqref{class2}
\begin{align}
n^{2}=&x_{0}\frac{\sqrt{6\mu} L^2}{Q}-\left(\sqrt{x_{0}}\right)^{3}\frac{\left(2/3\right)^{1/4} \mu ^{3/4} L^3}{Q^{5/2}}-\left(\sqrt{x_{0}}\right)^{4} \left(\frac{7 \mu  L^4}{3 Q^4}+\frac{3}{2}\right)+\left(\sqrt{x_{0}}\right)^{5}\frac{{\mu }^{1/4} L  \left(138 Q^4-439 \mu  L^4\right)}{ {{2}^{1/4}} 3^{3/4} 24Q^4 \left(Q^2\right)^{3/4}}\\
\omega=&\sqrt{x_{0}}\frac{{\left(2/3\right)}^{1/4} }{{\mu }^{1/4} L \sqrt{Q}}+x_{0}\frac{1}{2 Q^2}-\left(\sqrt{x_{0}}\right)^{3}\frac{ 6 Q^4-115 \mu  L^4}{ {2}^{1/4} 3^{3/4} 24 \mu ^{3/4} L^3 Q^{7/2}}-\left(\sqrt{x_{0}}\right)^{4}\frac{ 21 Q^4-197 \mu  L^4}{12 \sqrt{6\mu } L^2 Q^{5}}\\
\mathcal{A}=&\sqrt{x_{0}} \ 2^{3/4} {3}^{1/4} 2 \pi  {\mu }^{1/4} L \sqrt{Q} -x_{0}\frac{ \pi  \sqrt{6\mu } L^2}{Q}+\left(\sqrt{x_{0}}\right)^{3}\frac{\pi  \left(6 Q^4-97 \mu  L^4\right)}{6\ 2^{3/4} {3}^{1/4} {\mu }^{1/4} L Q^{5/2}}+ \left(\sqrt{x_{0}}\right)^4\ 3 \pi \left(1-\frac{8 \mu  L^4}{Q^4}\right)\\
S=&\sqrt{x_{0}} \ \frac{2^{3/4} {3}^{1/4} \pi \ {\mu }^{1/4} L \sqrt{Q}}{G}+x_{0} \ \frac{\pi  \sqrt{\mu/6 } L^2 }{ G Q}+\left(\sqrt{x_{0}}\right)^{3}\frac{\pi \left(47 \mu  L^4+6 Q^4\right)}{12\ 2^{3/4} {3}^{1/4} G {\mu }^{1/4} L Q^{5/2}}\\ \notag
&+\left(\sqrt{x_{0}}\right)^4 \ \frac{\pi \left(136 \left(\mu  L^4/Q^4\right)-15\right)}{18 G}\\
S/S_{\rm BH}=&2+\sqrt{x_{0}} \ \frac{2 \left(2/3\right)^{3/4} {\mu }^{1/4} L }{Q^{3/2}}+x_{0} \ \frac{7 \sqrt{2\mu/3} L^2}{Q^{3}}+\left(\sqrt{x_{0}}\right)^{3}\frac{ 103 \mu  L^4-10 Q^4}{2\ 2^{3/4} {3}^{1/4} {\mu }^{1/4} L Q^{9/2}}\,.
\end{align}

\section{Solutions for the $g_{k}(y)$}\label{app:sol}

Expanding $g(y)$ as in \eqref{expansiong} and choosing the parameter configuration of the AdS$_{2}\times\mathbb{S}^{2}$ branch, we see that the solutions for all $g_k(y)$ that satisfy the boundary conditions are polynomial in $y$. The first terms read as follows

\begin{align}
\frac{g_{0}(y)}{1-y^2}=&\frac{1}{Q^{2}}\\ 
\frac{g_{1}(y)}{1-y^{2}}=&\frac{-Q^{8} \left(y^{2}+1\right)+Q^{4} \mu  L^{4} \left(16 y^{2}+3\right)-9 \mu ^{2} L^{8}}{Q^{12}-9 Q^{8} \mu  L^{4}}\\ 
\frac{g_{2}(y)}{1-y^{2}}=&\frac{Q^{24} \left(y^2+1\right)^2+Q^{20} \mu  L^4 \left(-232 y^4+39 y^2-27\right)+Q^{16} \mu ^2 L^8 \left(6555 y^4-2218 y^2-592\right)}{Q^{14} \left(Q^4-30 \mu  L^4\right) \left(Q^4-9 \mu  L^4\right)^3}\\ \notag
&+\frac{3 Q^{12} \mu ^3 L^{12} \left(-21412 y^4+9939 y^2+3798\right)+27 Q^8 \mu ^4 L^{16} \left(7768 y^4-4248 y^2-3085\right)}{Q^{14} \left(Q^4-30 \mu  L^4\right) \left(Q^4-9 \mu  L^4\right)^3} \\ \notag
&+\frac{-4617 Q^4 \mu ^5 L^{20} \left(20 y^2-77\right)-240570 \mu ^6 L^{24}}{Q^{14} \left(Q^4-30 \mu  L^4\right) \left(Q^4-9 \mu  L^4\right)^3}\\
\frac{g_{3}(y)}{1-y^{2}}=&\frac{-Q^{44} \left(y^2+1\right)^3+Q^{40} \mu  L^4 \left(1273 y^6+87 y^4-145 y^2-15\right)}{Q^{20} \left(Q^4-63 \mu  L^4\right) \left(Q^4-30 \mu  L^4\right)^2 \left(Q^4-9 \mu  L^4\right)^5}\\ \notag
&+\frac{2 Q^{36} \mu ^2 L^8 \left(-85596 y^6+26710 y^4+15907 y^2+11113\right)}{Q^{20} \left(Q^4-63 \mu  L^4\right) \left(Q^4-30 \mu  L^4\right)^2 \left(Q^4-9 \mu  L^4\right)^5}\\ \notag
&+\frac{Q^{32} \mu ^3 L^{12} \left(8837031 y^6-4893135 y^4-730200 y^2-1854008\right)}{Q^{20} \left(Q^4-63 \mu  L^4\right) \left(Q^4-30 \mu  L^4\right)^2 \left(Q^4-9 \mu  L^4\right)^5}\\ \notag
&+\frac{-6 Q^{28} \mu ^4 L^{16} \left(37929171 y^6-27161595 y^4+80490 y^2-10643708\right)}{Q^{20} \left(Q^4-63 \mu  L^4\right) \left(Q^4-30 \mu  L^4\right)^2 \left(Q^4-9 \mu  L^4\right)^5}\\ \notag
&+\frac{18 Q^{24} \mu ^5 L^{20} \left(180123972 y^6-151235249 y^4+10041658 y^2-65511537\right)}{Q^{20} \left(Q^4-63 \mu  L^4\right) \left(Q^4-30 \mu  L^4\right)^2 \left(Q^4-9 \mu  L^4\right)^5}\\ \notag
&+\frac{-81 Q^{20} \mu ^6 L^{24} \left(321947241 y^6-302005267 y^4+27567259 y^2-165642213\right)}{Q^{20} \left(Q^4-63 \mu  L^4\right) \left(Q^4-30 \mu  L^4\right)^2 \left(Q^4-9 \mu  L^4\right)^5}\\ \notag
&+\frac{729 Q^{16} \mu ^7 L^{28} \left(152540040 y^6-152997410 y^4+9998425 y^2-132397703\right)}{Q^{20} \left(Q^4-63 \mu  L^4\right) \left(Q^4-30 \mu  L^4\right)^2 \left(Q^4-9 \mu  L^4\right)^5}\\ \notag
&+\frac{-13122 Q^{12} \mu ^8 L^{32} \left(14926950 y^6-14295073 y^4-3145538 y^2-31632593\right)}{Q^{20} \left(Q^4-63 \mu  L^4\right) \left(Q^4-30 \mu  L^4\right)^2 \left(Q^4-9 \mu  L^4\right)^5}\\ \notag
&+\frac{708588 Q^8 \mu ^9 L^{36} \left(130865 y^4-373185 y^2-1354486\right)+127545840 Q^4 \mu ^{10} L^{40} \left(875 y^2+8112\right)}{Q^{20} \left(Q^4-63 \mu  L^4\right) \left(Q^4-30 \mu  L^4\right)^2 \left(Q^4-9 \mu  L^4\right)^5}\\ \notag
&+\frac{-348200143200 \mu ^{11} L^{44}}{Q^{20} \left(Q^4-63 \mu  L^4\right) \left(Q^4-30 \mu  L^4\right)^2 \left(Q^4-9 \mu  L^4\right)^5}\\ \notag
\end{align}

\section{Adding a cosmological constant $\Lambda$}\label{app:lambda}
For the sake of completeness, here we shall comment on the case of a non-vanishing cosmological constant, $\Lambda\neq 0$. The regularity constraints, analogue to \eqref{eq:constr1}, for a non-vanishing $\Lambda$ read
\begin{align}\label{constraint1}
0=-n^2+x_0^2+\frac{Q^{2}\omega ^2 \left(n^2+x_0^2\right)^2}{x_0^2}-\mu  L^4 \omega ^2 (2 x_0 \omega +3)+\frac{\Lambda}{3}   \left(-3 n^4+6 n^2 x_0^2+x_0^4\right),
\end{align}
and
\begin{align}\label{constraint2}
0=\left(n^2+x_0^2\right)\left(\omega  \left(n^2+x_0^2\right)-x_0\right)+\mu  L^4 \omega^2\left( \omega \left(x_0^2-5 n^2\right)+3  x_0 \right)-\frac{\Lambda}{3}x_{0}\left(n^2+x_0^2\right)  \left(6  n^2  +2 x_0^2\right).
\end{align}
We shall focus on the neighbourhood of solutions for which $\omega=0$. Such solutions are non-rotating if $x_{0}=0$ in such a way that $\lim_{\omega\rightarrow 0}\omega/x_{0}\ne0$, while $\lim_{\omega\rightarrow 0} x_0/n=0$. On the other hand, if $\lim_{\omega\rightarrow 0}\omega/x_{0}=0$ and $\lim_{\omega\rightarrow 0}x_0/n\ne0$, the solutions correspond to an ultra-spinning limit and exhibit a non-compact horizon --- this only happens for $\Lambda<0$. Let us first consider the slowly rotating case. Imposing $\omega=0$, one solution is $x_{0}=0$ and $n^{2}=\left(1-4 \Lambda  Q^2-\sqrt{1-4 \Lambda  Q^2}\right)/\left(-2 \Lambda  \left(1-4 \Lambda  Q^2\right)\right)$. Then, in a neighbourhood of this solution, $\omega$ and $n^2$ read, in powers of $x_{0}$,

\begin{align}
n^{2}=&\frac{1-4 \Lambda  Q^2-\sqrt{1-4 \Lambda  Q^2}}{-2 \Lambda  \left(1-4 \Lambda  Q^2\right)}\\ \notag
&+x_0^2\Biggl(\frac{ Q^4 \left(72 \Lambda ^2 \mu  L^4-6\right)-6 \Lambda  \mu  L^4 Q^2 \left(7 \sqrt{1-4 \Lambda  Q^2}+6\right)-3 \mu  L^4 \left(\sqrt{1-4 \Lambda  Q^2}+1\right)-8 \Lambda  Q^6}{6 Q^4 \left(4 \Lambda  Q^2-1\right)}\Biggr)\\ \notag
&+\mathcal{O}\left(x_{0}^4\right)\\ 
\omega=&x_{0}\frac{1+ \sqrt{1-4 \Lambda  Q^2}}{2 Q^2}\\ \notag
&+x_{0}^{3}\Lambda ^2\frac{6 \Lambda  \mu  L^4 Q^2 \left(7 \sqrt{1-4 \Lambda  Q^2}+8\right)-3 \mu  L^4 \left(\sqrt{1-4 \Lambda  Q^2}+1\right)+4 Q^4 \left(-36 \Lambda ^2 \mu  L^4+2 \sqrt{1-4 \Lambda  Q^2}+1\right)-16 \Lambda  Q^6}{3 Q^4 \left(2 \Lambda  Q^2 \left(\sqrt{1-4 \Lambda  Q^2}-2\right)-\sqrt{1-4 \Lambda  Q^2}+1\right)}\\ \notag
&+\mathcal{O}\left(x_{0}^5\right).
\end{align}
On the other hand, performing an expansion of $g(y)$ around $x_{0}=0$ as in \eqref{expansiong}, the first non-vanishing term reads
\begin{align}\label{g0}
\frac{g(y)}{1-y^2}=&\frac{-\Lambda  \left(2 \Lambda  Q^2+\sqrt{1-4 \Lambda  Q^2}-1\right)}{\Lambda  Q^2 \left(\sqrt{1-4 \Lambda  Q^2}-3\right)-\sqrt{1-4 \Lambda  Q^2}+1}+\mathcal{O}\left(x_{0}^2\right)\, .
\end{align}
Thus, we conclude that AdS$_{2}\times\mathbb{S}^{2}$ is not corrected by ECG also when $\Lambda\neq 0$. However, it admits smooth corrections when the spin is turned on. This is analogous to the AdS$_{2}\times\mathbb{S}^{2}$ branch for $\Lambda=0$ discussed above and, in fact, taking the limit $\Lambda\rightarrow0$, the RHS of \eqref{g0} goes to $1/Q^{2}$ as expected. The expansions for the area and the entropy are
\begin{align}
\mathcal{A}=&\frac{8 \pi  Q^2}{\sqrt{1-4 \Lambda  Q^2}+1}\\ \notag
&+x_{0}^{2}\frac{2 \pi   \left(6 \Lambda  \mu  L^4 Q^2 \left(7 \sqrt{1-4 \Lambda  Q^2}+8\right)-3 \mu  L^4 \left(\sqrt{1-4 \Lambda  Q^2}+1\right)+4 Q^4 \left(-36 \Lambda ^2 \mu  L^4+2 \sqrt{1-4 \Lambda  Q^2}+1\right)-16 \Lambda  Q^6\right)}{3 Q^4 \sqrt{1-4 \Lambda  Q^2}}\\ \notag
&+\mathcal{O}\left(x_{0}^{4}\right)\\ 
S=&\frac{2 \pi  Q^2}{G (\sqrt{1-4 \Lambda  Q^2}+1)}-x_{0}^{2}\frac{2 \pi \left(Q^2 \left(18 \Lambda ^2 \mu  L^4-\sqrt{1-4 \Lambda  Q^2}+1\right)+3 \Lambda  \mu  L^4 \left(\sqrt{1-4 \Lambda  Q^2}+1\right)+4 \Lambda  Q^4\right)}{3 G Q^2 \left(\sqrt{1-4 \Lambda  Q^2}-1\right)}+\mathcal{O}(x_{0}^{4})\, .
\end{align}
Let us now focus our attention on the ultra-spinning case. When $\omega=0$ another solution is $x_{0}=\sqrt{3}/(2 \sqrt{-\Lambda })$ and $n^{2}=-1/4\Lambda$, which of course is only valid for a negative cosmological constant, $\Lambda<0$. For simplicity, we shall restrict to the neutral case $Q=0$. In a neighbourhood of this solution, $\omega$ and $n$ read, in powers of $x_{0}-\sqrt{3}/(2 \sqrt{-\Lambda })$,
\begin{align}\label{Sentropic1}
n^2=&-\frac{1}{4 \Lambda }+\frac{1}{2} \left(x_0-\frac{\sqrt{3}}{2 \sqrt{-\Lambda }}\right)^2 \left(-3 \Lambda ^2 \mu  L^4-1\right)+\frac{1}{6} \left(x_0-\frac{\sqrt{3}}{2 \sqrt{-\Lambda }}\right)^3 \left(\sqrt{3} \sqrt{-\Lambda }+21 \sqrt{3} \Lambda ^2 \sqrt{-\Lambda } \mu  L^4\right)\\ \notag
&+\mathcal{O}\left(\left(x_0-\frac{\sqrt{3}}{2 \sqrt{-\Lambda }}\right)^4\right)\\ \label{Sentropic2}
\omega=&\Lambda  \left(x_0-\frac{\sqrt{3}}{2 \sqrt{-\Lambda }}\right)-\frac{1}{2} \sqrt{3} \sqrt{-\Lambda } \Lambda  \left(x_0-\frac{\sqrt{3}}{2 \sqrt{-\Lambda }}\right)^2-\frac{1}{6} 7 \Lambda ^2 \left(x_0-\frac{\sqrt{3}}{2 \sqrt{-\Lambda }}\right)^3 \left(3 \Lambda ^2 \mu  L^4+1\right)\\ \notag
&+\mathcal{O}\left(\left(x_0-\frac{\sqrt{3}}{2 \sqrt{-\Lambda }}\right)^4\right)\,.
\end{align}
On the other hand, the area and the entropy are
\begin{align}
\mathcal{A}=&-\frac{2 \sqrt{3} \pi }{(-\Lambda )^{3/2}\left(x_{0}-\frac{\sqrt{3}}{2 \sqrt{-\Lambda }}\right)}+\frac{7 \pi }{\Lambda }+\frac{7 \pi \left(6 \Lambda ^2 \mu  L^4-1\right)}{2 \sqrt{3} \sqrt{-\Lambda }} \left(x_0-\frac{\sqrt{3}}{2 \sqrt{-\Lambda }}\right)+\mathcal{O}\left(\left(x_0-\frac{\sqrt{3}}{2 \sqrt{-\Lambda }}\right)^{2}\right)\\ 
S=&-\frac{\pi  \sqrt{3}}{2G(-\Lambda )^{3/2}\left(x_0-\frac{\sqrt{3}}{2 \sqrt{-\Lambda }}\right)}+\frac{7 \pi }{4 G \Lambda }+\frac{\pi \left(6 \Lambda ^2 \mu  L^4-7\right)}{8 \sqrt{3} G \sqrt{-\Lambda }}\left(x_0-\frac{\sqrt{3}}{2 \sqrt{-\Lambda }}\right)+\mathcal{O}\left(\left(x_0-\frac{\sqrt{3}}{2 \sqrt{-\Lambda }}\right)^{2}\right).
\end{align}
The latter pair of quantities are only well defined on an anular neighbourhood centred at $x_0=\sqrt{3}/2 \sqrt{-\Lambda }$. However, this does not mean that there is no solution when $x_0=\sqrt{3}/2 \sqrt{-\Lambda }$. Let us recall that, as long as $\omega/x_0$ remains finite, so does $g'(1)$, according to \eqref{conditions}. Then, the coordinates can be chosen to parametrize a manifold of topology AdS$_{2}\times \mathbb{S}^{2}$ by identifying canonically the coordinate $\phi$, \textit{i.e.} $\phi\sim\phi+2\pi$ (see equation \eqref{solmassaged}), and the metric becomes regular everywhere. However, if $\omega/x_0=0$ then $g'(1)$ also vanishes and the metric does not describe a regular geometry on AdS$_{2}\times \mathbb{S}^{2}$. Thus, in order to obtain a solution also at the parameter configuration
\begin{equation}\label{Sentropic3}
x_{0}=\sqrt{3}/2 \sqrt{-\Lambda},\,\,\,\,\,\omega=0,\,\,\,\,\, n^{2}=-1/4\Lambda,
\end{equation}
let us rewrite  \eqref{solmassaged} in terms of a new angular coordinate
\begin{equation}
\varphi=\frac{x_{0}}{\omega}\phi
\end{equation} 
and identify it with arbitrary period, $\varphi\sim\varphi+\Delta \varphi$. The equations \eqref{constraint1} and \eqref{constraint2} are unchanged by this coordinate transformation, so \eqref{Sentropic3} constitute a solution. Since $g'(1)=0$, the topology our coordinates parametrize is that of AdS$_{2}\times \mathbb{S}^{1}\times\mathbb{R}$, and the metric is regular everywhere. The horizon has become non-compact, with topology $\mathbb{S}^{1}\times\mathbb{R}$, and is infinitely large, in the sense that the proper length of coordinate curves tangent to $\partial_{y}$ (which extend from $y=-1$ to $y=1$) is infinite. However, the horizon has a finite area, $\mathcal{A}=2\Delta\varphi$, and Wald's correction to the Bekenstein--Hawking entropy vanishes, as can be deduced from \eqref{WaldCorrection}, because now both $g(1)$ and $g'(1)$ are zero. Nevertheless, one can check that the profile of the solution is not going to be the same as in Einstein gravity. This solution of ECG is analogous to the super-entropic black holes of Ref.~\cite{Hennigar:2014cfa}, in the sense that both have non-compact horizons with finite area and can be understood as an entropy-divergent limit of a rotating solution.  Thus, it would be interesting to study whether this solutions do or do not respect the Isoperimetric Inequality in the context of extended black hole thermodynamics. However, further investigation in these lines is left for future work.

\bibliographystyle{apsrev4-1} 
\vspace{1cm}
\bibliography{Gravities} 

\end{document}